\documentclass[twocolumn,times]{aastex61}
\pdfoutput=1 
\usepackage{amsmath,amstext,natbib,float,epsfig,multirow,soul}
\usepackage[figure,figure*]{hypcap}

\newcommand{\lyaf}[1]{Ly$\alpha$ forest}
\newcommand{\lya}[1]{Ly$\alpha$}

\newcommand{\hMpc}{\ensuremath{h^{-1}\,\mathrm{Mpc}}}

\newcommand{\hMpccube}{\ensuremath{\, h^{-3}\,\mathrm{Mpc}^3 }}

\newcommand{\dperp}{\ensuremath{\langle d_{\perp} \rangle}}

\shorttitle{$z\sim 2$ Cosmic Voids from IGM Tomography}
\shortauthors{Krolewski et al.}

\begin{document}

\title{A Detection of $\lowercase{z}\sim 2.3$ Cosmic Voids from
3D Lyman-$\alpha$ Forest Tomography in the COSMOS Field}

\author{Alex Krolewski}
\affiliation{Department of Astronomy, University of California at Berkeley, New Campbell Hall, Berkeley, CA 94720, USA}

\author{Khee-Gan Lee}
\altaffiliation{Hubble Fellow}
\affiliation{Lawrence Berkeley National Lab, 1 Cyclotron Road, Berkeley, CA 94720, USA}

\author{Martin White}
\affiliation{Department of Astronomy, University of California at Berkeley, New Campbell Hall, Berkeley, CA 94720, USA}
\affiliation{Lawrence Berkeley National Lab, 1 Cyclotron Road, Berkeley, CA 94720, USA}

\author{Joseph F.\ Hennawi}
\affiliation{Department of Physics, Broida Hall, University of California at Santa Barbara, Santa Barbara, CA 93106, USA}

\author{David J.\ Schlegel} 
\affiliation{Lawrence Berkeley National Lab, 1 Cyclotron Road, Berkeley, CA 94720, USA}

\author{Peter E.\ Nugent}
\affiliation{Lawrence Berkeley National Lab, 1 Cyclotron Road, Berkeley, CA 94720, USA}
\affiliation{Department of Astronomy, University of California at Berkeley, New Campbell Hall, Berkeley, CA 94720, USA}

\author{Zarija Luki\'c}
\affiliation{Lawrence Berkeley National Lab, 1 Cyclotron Road, Berkeley, CA 94720, USA}

\author{Casey W.\ Stark}
\affiliation{Google Inc., 1600 Amphitheatre Parkway, Mountain View, CA 94043, USA}

\author{Anton M.\ Koekemoer}
\affiliation{Space Telescope Science Institute, 3700 San Martin Drive,
Baltimore MD 21218, USA}

\author{Olivier Le F\`evre}
\affiliation{Aix Marseille Universit\'e, CNRS, LAM (Laboratoire d'Astrophysique  de Marseille) UMR 7326, 13388, Marseille, France}

\author{Brian C.\ Lemaux}
\affiliation{Department of Physics, University of California, Davis, One Shields Ave., Davis, CA 95616, USA}
\affiliation{Aix Marseille Universit\'e, CNRS, LAM (Laboratoire d'Astrophysique  de Marseille) UMR 7326, 13388, Marseille, France}

\author{Christian Maier}
\affiliation{University of Vienna, Department of Astrophysics, Tuerkenschanzstrasse 17, 1180 Vienna, Austria}

\author{R.\ Michael Rich}
\affiliation{Department of Physics and Astronomy, University of California at Los Angeles, Los Angeles, CA 90095, USA}

\author{Mara Salvato}
\affiliation{Max Planck Institute for Extraterrestrial Physics, Gie{\ss}enbachstraße 1, 85741 Garching bei M\"unchen, Germany}

\author{Lidia Tasca}
\affiliation{Aix Marseille Universit\'e, CNRS, LAM (Laboratoire d'Astrophysique  de Marseille) UMR 7326, 13388, Marseille, France}

\correspondingauthor{Alex Krolewski}
\email{krolewski@berkeley.edu}

\begin{abstract}
We present the most distant detection of cosmic
voids ($z \sim 2.3$) and the first detection
of three-dimensional voids in the Lyman-$\alpha$
forest. We used a 3D tomographic map of the absorption with effective comoving spatial resolution
of $2.5\,h^{-1}\mathrm{Mpc}$ and volume of $3.15\times 10^5\,h^{-3}\mathrm{Mpc}^3$, which was
reconstructed from moderate-resolution Keck-I/LRIS spectra of 240 background
Lyman-break galaxies and quasars in a $0.16\,\mathrm{deg}^2$ footprint in the COSMOS field.  Voids were detected using a spherical overdensity
finder calibrated from hydrodynamical simulations of the intergalactic medium. This allows us to identify voids in the IGM corresponding to voids in the underlying matter density field, yielding a consistent
volume fraction of voids in both data (19.5\%) and simulations (18.2\%).
We fit excursion set models to the void radius function and compare the radially-averaged stacked profiles of large voids
($r > 5$ \hMpc{}) to stacked voids in mock observations and the simulated density field.
Comparing with 432 coeval galaxies with spectroscopic redshifts in the same volume as the tomographic map, we find that the
tomography-identified voids are underdense in galaxies
by 5.95$\sigma$ compared to random cells.
\end{abstract}

\section{Introduction}
Cosmic voids offer a laboratory for studying cosmology and galaxy formation in extreme environments.  Voids are large (Mpcs to tens of Mpcs), slightly prolate regions nearly devoid of galaxies, which constitute the majority of the universe's volume \citep{vp+11}.  Voids are surrounded by the beaded,
filamentary network of the cosmic web and expand and evacuate as matter streams onto filaments and collapses
into halos \citep{bond+96}.  Matter streams outward most quickly in the center of voids, where the density is lowest, creating a so-called bucket profile with a uniform inner density \citep[$\delta \sim -0.7-0.9$; ][]{ham+14,sut+14}. 
The exact shape of the profile is dependent on both the void finder and the large-scale
environment of the void under consideration: small voids are often subvoids within a large-scale
overdensity and are surrounded by a ridge of higher density, while large voids \citep[as well as voids found by spherical overdensity finders; see][]{white_padmanabhan:2017} typically
have a smooth profile approaching the mean density from below \citep{ham+14,cai:2016}.
While isolated voids become more isotropic over time \citep{sheth_vdw:2004},
voids in the real universe remain prolate due to external tides and collisions with neighboring
sheets and filaments \citep{vp+11}.

Voids are especially useful for studying components of the universe that cluster weakly,
such as dark energy \citep{leepark+09,lavwand+12}
 or massive neutrinos \citep{vn+13,mass+15,bd+16}: since voids are underdense in the clustered
components of the universe (dark matter and baryons), unclustered components will have a maximal effect on the dynamics within
voids \citep{gv+04}.  Voids are also sensitive probes of modified gravity theories,
which may be screened in higher density regions \citep{clamp+13}.

Prospects for void cosmology have been studied using several different observables.  Since voids
are spherical on average, the Alcock-Paczynski test \citep{ap:1979} can be performed on sufficiently large stacks of voids \citep{ryd95,lavwand+12}.  Other sensitive observables include 
void-galaxy cross-correlations and redshift-space distortions \citep{cai:2016,ham+17};
the integrated Sachs-Wolfe effect from stacked voids \citep{granett:2008,cai:2017,kovacs:2017};
weak lensing of stacked voids \citep{higuchi:2013,krause:2013,melchior:2014,clampitt_jain:2015,barreira:2015,gruen:2016,cai:2017};
void counts to probe modified gravity \citep{li:2012,clamp+13,lam:2015,cai:2015,zivick:2015}
or dark energy
\citep{pis+15,pollina:2016}; and
void ellipticities \citep{parklee07,bos:2012}.
Extending the study of cosmic voids to higher redshifts
could allow for better constraints on redshift-dependent models,
such as early dark energy \citep{dor_rob:2006}.

Studying galaxies in voids can illuminate the influence of environment on galaxy
evolution. $N$-body simulations show that the halo mass function abruptly changes from sheets
to voids, leading to a dearth of dwarf galaxies in voids.  This is the so-called ``void phenomenon,''
originally identified as a tension with $\Lambda$CDM by \citet{peebles:2001} but explained
in the context of the halo model by \citet{tc09}.
Comparisons of void galaxies to galaxies in average environments suggest that the change in the stellar
mass function plays a dominant role in modifying galaxy properties as compared to the field
\citep{hoyle+05,tinker:2008,alp15,penny:2015,beygu:2016} and void galaxies show a similar diversity
in morphology to field galaxies of the same stellar mass \citep{beygu+17}.
Recently some hints have emerged that void galaxies may have a slightly higher mass-to-light
ratio than field galaxies of the same mass \citep{alp15}, slightly higher HI masses
at low stellar mass \citep{beygu:2016} and slightly enhanced star formation rate to HI mass ratio \citep{kreckel:2012}, although these effects remain quite subtle.  Since the global star formation
rate of the universe is much higher at $z \sim 2$ at $z \sim 0$, it would be interesting to study
whether stellar mass remains the primary driver of void galaxy properties at $z \sim 2$,
or whether environment begins to play a more significant role.

Observational studies of voids have been limited to low to moderate
redshift where sufficiently dense galaxy surveys are available
to identify voids.  Voids have been identified in 2dF \citep{cecc+06},
SDSS \citep{pan:2012,sut+12}, VIPERS \citep{mich+14}, BOSS \citep{mao+17}, DES \citep{san+17}, and
DEEP2 \citep{conroy+05}.  The SDSS and BOSS voids have also been used for cosmological analyses \citep{sutter:2014,ham+16,mao+17b}.  Finding voids with radius of a few Mpc requires a large-volume galaxy survey with resolution of a few Mpc, which becomes
increasingly difficult above $z \sim 1$ \citep{stark:2015a}.

At higher redshifts, Lyman-$\alpha$ forest tomography \citep{pichon2001,cau+08,lee_obs_req}
offers an alternative method for obtaining large-volume, densely
spaced surveys of the matter density field.  Using spectroscopic observations
of closely-spaced quasars and Lyman-break galaxies, Lyman-$\alpha$
forest tomography can reconstruct the 3D intergalactic medium absorption field with
resolution of a few Mpc and on cosmological volumes of $10^6$ $h^{-3}$ Mpc$^{3}$ \citep{lee_obs_req}.  This technique allows for recovery of the cosmic
web with comparable fidelity to $z < 0.5$ galaxy surveys \citep{lee_white+16,krolewski:2017},
which requires considerably greater spatial resolution than $z \sim 2$ galaxy surveys
can provide.  At $z \sim 2.5$, absorption with optical depth unity
arises from neutral hydrogen with three times the mean density; thus,
the Lyman $\alpha$ forest is ideal for probing underdense structures
such as voids.  Indeed, \citet{stark:2015a} found that a simple spherical overdensity
void finder could recover $r \geq 6$ \hMpc{} voids in the IGM flux field at 60\%
fidelity, allowing detection of $\sim 100$ such voids in a 1 deg$^2$ survey.

In this paper, we make the first detection of $z \sim 2$
cosmic voids in the 3D Lyman-$\alpha$ forest using the COSMOS Lyman-Alpha Mapping and Tomography Observations (CLAMATO) survey \citep{lee+17}.  CLAMATO is the first
survey to systematically use Lyman-break galaxies for Lyman-alpha
forest analysis.  It has produced a 3D map of the IGM absorption field 
with resolution $2.5$ \hMpc{} and volume $3.15 \times 10^5$ h$^{-3}$ Mpc$^3$, using Keck-I/LRIS observations
of the central $0.16$ deg$^2$ of the COSMOS field.  


While we are not the first to consider voids in the IGM, this work is distinct
from previous observational efforts: \citet{tejos+12} worked at $z \sim 0$;
\citet{roll+03} used only four sightlines, leading to large uncertainties; and \citet{viel+08} were limited
to analyzing flux in 1D skewers.

The detection of $z \sim 2$ voids extends observational studies of voids
to a much higher redshift range.  
In the future, high-redshift voids
could allow for studies of the
redshift evolution of void galaxies and void properties
over a much larger redshift baseline, and 
better constraints on redshift-dependent
dark energy and modified gravity models.

We begin by describing the data (Section~\ref{sec:data}) and simulations (Section~\ref{sec:sims})
used in this paper.  Next we determine appropriate spherical overdensity thresholds by matching
the void fraction in mock tomographic observations to the fraction of true voids
in the density field
(Section~\ref{sec:calib_vf}).  We apply these thresholds
to data in Section~\ref{sec:app_to_data}.  In Section~\ref{sec:void_galaxy_cic}, we compare the tomography-identifed voids to the positions of 
coeval galaxies with spectroscopic redshifts,
and find that the voids are $\sim 6\sigma$ underdense in coeval galaxies.
We discuss the properties of the voids in Section~\ref{sec:void_prop} (including the void radius function
and stacked void profile)
and present our conclusions in Section~\ref{sec:conclusions}.

In this paper we use a flat $\Lambda$CDM cosmology with $\Omega_m = 0.31$ and $h = 0.7$.  While the simulations use a slightly different
cosmology (see Section~\ref{sec:sims}), the differences are small
enough that the discrepancy
will have negligible impact on the results presented here.

\vspace{500pt}

\section{Data}
\label{sec:data}
\label{sec:lya}
We identify voids in the reconstructed IGM tomographic map from the first
data release of the CLAMATO survey\footnote{We use CLAMATO v4,
available here:
\url{https://doi.org/10.5281/zenodo.1292459}.}. The observations are described in detail 
by \citet{lee+17}, but we briefly summarize the pertinent details here.

The survey targeted $2.3 < z < 3$ background Lyman-break galaxies and quasars with the LRIS spectrograph \citep{oke:1995,steidel:2004} on the Keck-I telescope at Maunakea, Hawai'i,
to measure
the foreground Lyman-$\alpha$ forest absorption. This program targeted the COSMOS field
to take advantage
of rich existing datasets and achieve a high targeting
efficiency.
We observed 23 slitmasks (18 regular slitmasks and 5 ``special'' slitmasks
designed to fill in gaps in coverage) with $\sim 20$ targets per mask.
We successfully reduced 437 galaxies and AGN, of which 289 had high-confidence
redshifts, and 240 were usable for the Lyman-$\alpha$ forest analysis
at our targeted absorption redshift range of $2.05<z<2.55$. The primary
criterion for the selection of the background spectra was
the signal-to-noise ratio on the continuum in the Lyman-$\alpha$ forest (i.e.\ ratio of estimated continuum to pixel noise; hereafter we refer to this quantity
as ``S/N''): we required
 S/N $\geq 1.2$ per pixel.

The intrinsic continua of the background sources were estimated using
mean-flux regulation \citep{lee+12,lee:2013}, which adjusts the mean Lyman-$\alpha$ forest transmission within each sightline to be consistent with $\langle F(z)\rangle$ estimates from the literature --- we used \citet{fg:2008}. Based on \citet{lee+12}, we estimate that the continuum errors are approximately $\sim 10\%$ rms 
for the noisiest spectra ($\mathrm{S/N} \sim 2$ per pixel) and improve to $\sim 4\%$ rms for $\mathrm{S/N} \sim 10$ spectra.

From the observed flux density and the fitted continuum,
we compute the \lya{} forest fluctuations, $\delta_F$:
\begin{equation}
\delta_F = \frac{f}{C \langle F(z) \rangle} - 1
\label{eqn:deltaf}
\end{equation}
where $\langle F(z) \rangle$ is the mean \lya{} transmission from
\citet{fg:2008} (the power-law fit from Table 5, including metals, with bins of width $\Delta z = 0.1$).

We use these values of $\delta_F$ as input for the Wiener-filter tomographic reconstruction.  To avoid a flared map geometry, we use a constant
conversion between redshift and comoving distance, ${d \chi}/{dz}$,
and a constant transverse comoving distance $\chi$, both
evaluated at $z = 2.3$.  
With a fixed angular footprint on the sky, 
this amounts to a $\sim 20$\% change in the reconstruction
kernel size over the length of the map.  While our mocks lack
this redshift-dependent reconstruction kernel,
we find that our results are virtually unchanged
when we use an evolving $\chi(z)$ and ${d \chi}/{dz}(z)$.
Specifically, the volume fraction of voids drops from 19.5\% to 19.2\%
(0.2$\sigma$), the voids remain $\sim 6 \sigma$
underdense in coeval galaxies,
and the void radius function and void profile change by $<1\sigma$
at all bins.
Thus, we keep the simpler redshift- and angle-distance conversions
presented above, but caution that future, more detailed analysis
will likely require more accurate coordinate conversions
and thus a more complex map geometry.

We define an output grid with cells of comoving
size 0.5 \hMpc{}, transverse dimensions 30 \hMpc{} $\times$ 24 \hMpc{},
and line-of-sight length 438 \hMpc{}, corresponding to $2.05 < z < 2.55$.
Thus the total comoving volume is $3.15 \times 10^5$ \hMpccube{} over an survey
geometry which is elongated along the line-of-sight (redshift) dimension but considerably smaller
across the transverse dimensions.
The effective sightline spacing varies along the line of sight from
2.22 \hMpc{} at $z = 2.25$ to
3.15 \hMpc{} at $z = 2.45$.

We use a Wiener filtering algorithm developed by \citet{stark_protoclusters}
to reconstruct the 3D IGM absorption field:
\begin{equation}
\delta_F^{\textrm{rec}} = \textbf{C}_{\textrm{MD}} \cdot
(\textbf{C}_{\textrm{DD}} + \textbf{N})^{-1} \cdot \delta_F
\label{eqn:wiener}
\end{equation}
where $\textbf{N}$ is the noise covariance, $\textbf{C}_{\textrm{DD}}$
is the data-data covariance, and $\textbf{C}_{\textrm{MD}}$ is the map-data
covariance. 
We assume that the noise covariance is diagonal, with $\textbf{N}_{ij}
= n_i^2 \delta_{ij}$ where $n_i$ is the pixel noise.  To avoid
weighting any sightlines too heavily, we set a minimum noise level
of 0.2.  We further assume
that $\textbf{C}_{\textrm{MD}} = \textbf{C}_{\textrm{DD}} = C$:
\begin{equation}
C = \sigma_F^2 \exp{\left[-\frac{\Delta x_{\perp}^2}{2 l_{\perp}^2}
- \frac{\Delta x_{\parallel}^2}{2 l_{\parallel}^2}
\right]}
\label{eqn:covariance}
\end{equation}
We use $\sigma_F^2 = 0.05$, $l_{\perp} = \dperp = 2.5$ \hMpc{}, and 
$l_{\parallel} = 2.0$ \hMpc{}.
While in previous works we have
additionally Gaussian-smoothed the output tomographic reconstruction, in this paper
we apply no additional smoothing to the map, following \citet{stark:2015a}.

Hereafter we identify voids in the Wiener-filtered map rather
than in the pixel-level data.  While it should be possible to develop
a void finder that can be applied
directly to the pixel-level data (a method which could in principle also be extended
to the sparsely and irregularly sampled galaxy field), we leave the development of this
method to future work.

\begin{figure}[htb!]
\includegraphics[width=0.5\textwidth,clip=true, trim=10 80 10 10]{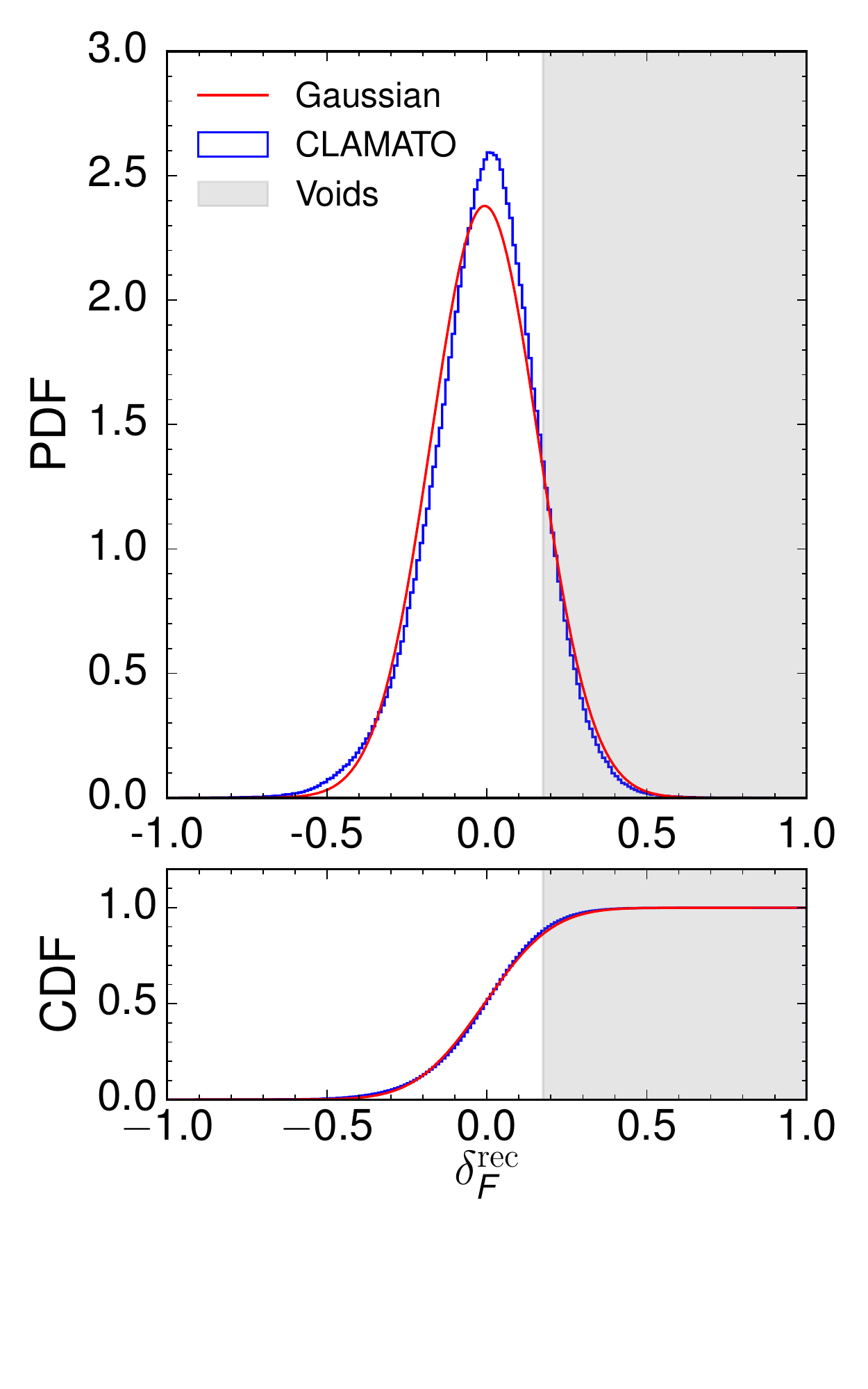}
\caption[]{\small Probability density and cumulative
distributions of $\delta_F^{\textrm{rec}}$ in the CLAMATO
map compared to a Gaussian.  The lower edge of the gray shaded
region is the threshold for average void density, $\delta_F^{\textrm{rec}} = 0.175$.
}
\label{fig:deltaf_hist}
\end{figure}

Figure~\ref{fig:deltaf_hist} shows the distribution
of $\delta_F^{\textrm{rec}}$ in the Wiener-filtered map
and overplots a Gaussian distribution with the same
mean and standard deviation.  Although the distribution
of $\delta_F^{\textrm{rec}}$ is reasonably well-approximated
by a Gaussian, particularly in the high $\delta_F^{\textrm{rec}}$ region where the voids lie,
the underlying density field smoothed on scales of 2.5
\hMpc{} is quite non-Gaussian, indicating that there
is cosmological information in the presence and
distribution of voids beyond the two-point statistics
in the map.

\vspace{20pt}

\section{Simulations}
\label{sec:sims}  

We use mock tomographic reconstructions from Ly$\alpha$ forest
simulations to both calibrate the thresholds for the spherical overdensity
void finder
and understand the effects of survey geometry and sample variance on our results. 
We use both hydrodynamic simulations of the IGM and $N$-body simulations of the density
field with the Ly$\alpha$ forest modeled using the fluctuating Gunn-Peterson approximation.
Each simulation has its advantages and disadvantages: the hydrodynamic simulation more accurately
models the physics of the IGM but is hampered by a relatively small volume of (100 \hMpc{})$^3$;
the larger (256 \hMpc{})$^3$ $N$-body simulation enables us to create many realizations of CLAMATO-like
volumes with approximately the correct survey geometry (though considerably shorter along the line of sight)
but its IGM prescription is only approximate.  Throughout this paper we use both simulations
and wherever possible we endeavor to compare the $N$-body and hydrodynamic simulation results
to ensure
robustness to different simulation methods and different included physics.

\vspace{10pt}

\subsection{Hydrodynamical Simulations}
The hydrodynamic simulations of the IGM are generated with the N-body plus Eulerian
hydrodynamics $\textsc{NYX}$ code \citep{alm_nyx}.  It has a 100 $h^{-1}$ Mpc box size with $4096^3$ cells and particles, resulting in a dark matter
particle mass of $1.02 \times 10^6 h^{-1} M_{\odot}$
and spatial resolution of 24 $h^{-1}$ kpc.  As discussed in \citet{lukic_nyx},
this resolution is sufficient to resolve the filtering scale below which
the IGM is pressure supported and to reproduce the $z=2.4$ flux statistics to percent accuracy within the range of physics included (we neglect radiative transfer
and do not model high column-density systems well).
We use a snapshot at $z = 2.4$.
This simulation uses a flat $\Lambda$CDM cosmology with $\Omega_m = 0.3$, $\Omega_b = 0.047$, $h = 0.685$, $n_s = 0.965$, and $\sigma_8 = 0.8$, consistent with latest Planck measurements \citep{pl16}.
It uses the ionizing background prescription of \citet{hm+96}, producing an IGM temperature-density relationship with $T_0 \sim 10^4$
K and  $\gamma \sim 1.55$ at $z = 2$.
This simulation does not model star-formation and hence has no feedback from stars, galaxies, or AGNs, 
but these are expected to have a negligible effect on the Ly$\alpha$ forest statistics \citep{viel:2013}.

We generate $512^2$ 
absorption skewers with a spacing of 0.2 $h^{-1}$ Mpc and
sample from these skewers to create mock data.  We compute the Ly$\alpha$
forest flux fluctuation along each skewer, then shift to redshift
space and Doppler broaden the skewers using the gas temperature.  The \ion{H}{1} optical depths,
$\tau$, in the mock spectra are adjusted to match the mean flux from
\citet{fg:2008} at $z = 2.3$ ($\langle F \rangle$ = 0.8189); we use a single mean flux throughout
the entire line of sight direction since neither simulation box is as
long as the line-of-sight length of the map.
Absorption skewers are randomly selected with mean sightline spacing $\langle d_{\perp} \rangle = 2.5$ \hMpc{} and rebinned along the line of sight
with resolution 0.84 $h^{-1}$ Mpc, corresponding to the 1.2 \AA\ LRIS pixels.
Using a single sightline spacing is approximate, 
as the mean transverse separation
of CLAMATO sightlines varies with redshift \citep{lee+17};
our choice of $\langle d_{\perp} \rangle = 2.5$ \hMpc{} is slightly conservative
compared to the CLAMATO $\langle d_{\perp} \rangle = 2.37$ \hMpc{}.
This difference should not be significant since we use the same 
correlation lengths for the tomographic reconstructions ($l_\perp$ and $l_\parallel$ in Equation~\ref{eqn:covariance}) in both mocks and
data.
Finally, the skewers are smoothed with a Gaussian kernel with 2.8 \hMpc{} FWHM
($\sim 4$ \AA) to account for the spectral resolution of LRIS at 4000 \AA.

\begin{figure}[htb!]
\includegraphics[width=0.5\textwidth,clip=true, trim=0 0 0 0]{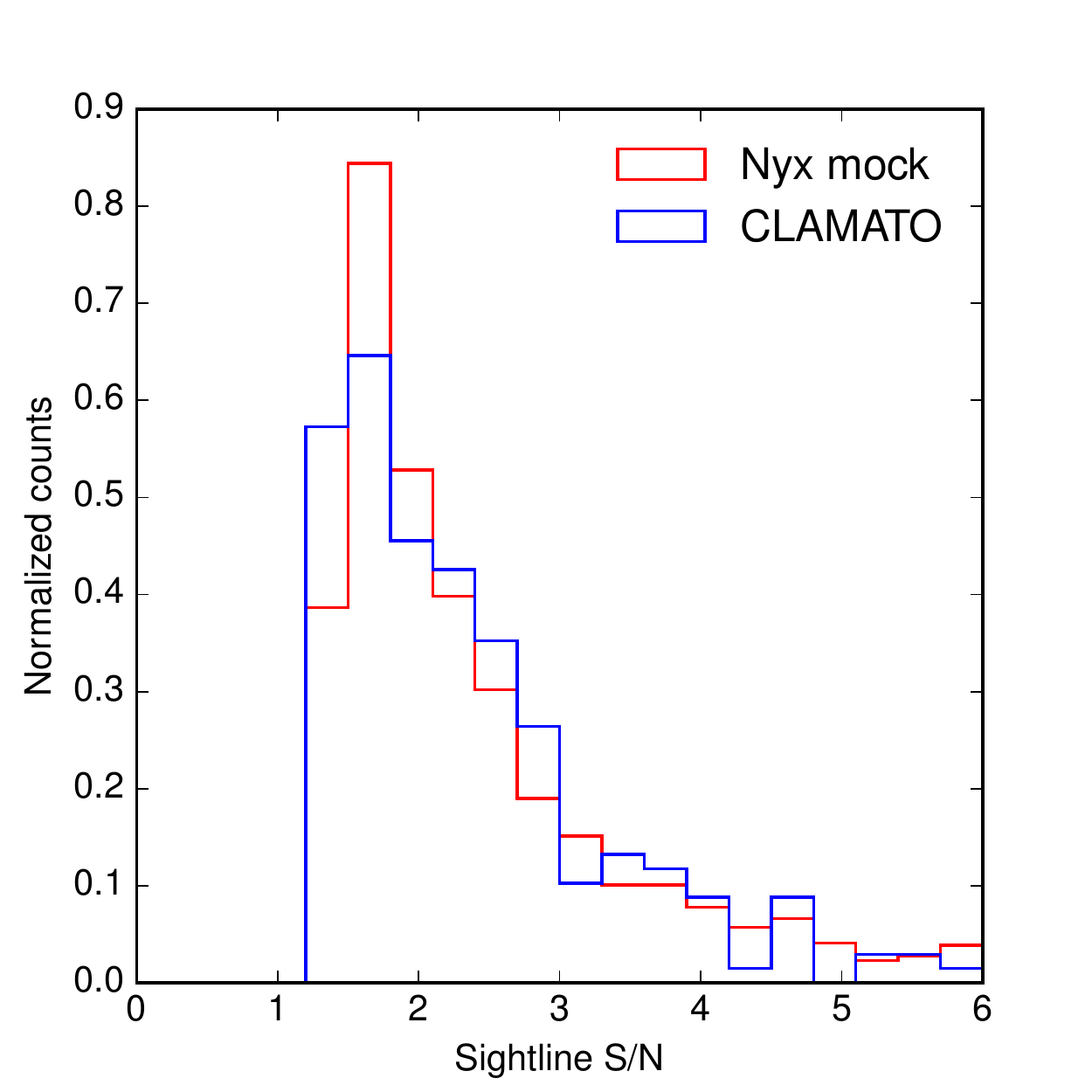}
\caption[]{\small Distribution of S/N per pixel for CLAMATO sightlines and Nyx mock sightlines.
}
\label{fig:pixel_noise}
\end{figure}

We add both random noise and correlated continuum error to each
skewer.
Random noise is simulated assuming the
S/N per pixel is a unique constant for each skewer.  To determine S/N for each skewer, we draw from a power-law S/N
distribution $\textrm{d}n_{\textrm{los}}/\textrm{dS/N}
\propto \textrm{S/N}^{-\alpha}$ \citep[][hereafter S15b]{stark_protoclusters},
where S/N ranges between 1.4 and infinity.
From S15b, we use $\alpha = 2.7$ for the $\langle d_{\perp} \rangle = 2.5$ \hMpc{} reconstructions.
The minimum S/N of 1.4 in the mock sightlines is slightly higher
than the minimum S/N of 1.2 in CLAMATO; the S/N distribution in CLAMATO rolls over below S/N of 1.5,
perhaps
owing to the difficulty of determining redshifts for low-S/N galaxies.
Therefore, a minimum S/N of 1.4 provides the best match to the
CLAMATO S/N distribution, with median S/N of 2.1 in CLAMATO
and 2.15 in the mock sightlines (Figure~\ref{fig:pixel_noise}).
We then use the S/N for each sightline to determine the
pixel noise $n$ (i.e. the error on $\delta_F$):
\begin{equation}
n = \frac{1}{\textrm{S/N}\langle F \rangle}
\label{eqn:signal_to_noise}
\end{equation}
Subsequently, we add a random Gaussian deviate with standard deviation
$n$
to the $\delta_F$ values in each pixel
and use the resulting noisy $\delta_F$ and $n$
as input to the Wiener filter (Equation~\ref{eqn:wiener}).

We also
model continuum-fitting error:
\begin{equation}
F_{\textrm{obs}} = \frac{F_{\textrm{sim}}}{1+\delta_{\textrm{cont}}}
\label{eqn:cont_err}
\end{equation}
where 
$\delta_{\textrm{cont}}$ is a random Gaussian
deviate, identical for all pixels within a skewer, with mean 0 and standard deviation $\sigma_{\textrm{cont}}$.
Following \citet{lee+12}, $\sigma_{\textrm{cont}}$ is a function
of the S/N, with lower S/N spectra having higher continuum error
and vice versa. We fit a function to the data points in Figure 8
of \citet{lee+12} for $z = 2.35$
\begin{equation}
\label{eqn:cont_err_sn}
\sigma_{\textrm{cont}} = \frac{a}{S/N} + b
\end{equation}
where $a = 0.2054$ is a free parameter fit to the data
and $b = 0.015$ is the rms fitting error in the absence of continuum
structure and noise, to which the continuum error should asymptote in
the case of infinite signal-to-noise.
To be conservative, we cap the continuum error for S/N $>10$ at 4\%.

We apply the same Wiener-filter interpolation to the mock
sightlines as to our data, with the same noise floor of 0.2 as in the data.
Just as in the data reconstruction, we use pixels 0.5 \hMpc{} on a side.

\subsection{Large-Volume N-body Simulations}
The hydrodynamical simulation is too small ($L=100\,\hMpc$) to mimic
the elongated CLAMATO survey geometry. 
To better understand the effect of survey geometry and sample variance on our
results,
we therefore also use a larger $N$-body
simulation \citep{white+10}. This is a publicly-available simulation used in our 
previous papers \citep{lee+15,stark:2015a,stark_protoclusters}, so we describe
it only briefly here.

The $N$-body simulation uses $2560^3$ particles of $8.6 \times 10^7$ $h^{-1} M_{\odot}$
in a 256 \hMpc{} periodic box.  The cosmological parameters are $\Omega_m = 0.31$,
$\Omega_b h^2 = 0.022$, $h = 0.677$, $n_s = 0.9611$, $\sigma_8 = 0.83$,
and initial conditions are generated using second-order Lagrangian perturbation theory
at $z _{ic} = 150$.  The particles were evolved forward using the TreePM code of
\citet{white:2002}, and we use output at $z = 2.5$.  The Ly$\alpha$ absorption field
was generated with the fluctuating Gunn-Peterson approximation assuming a pressure filtering scale of 100 $h^{-1}$ kpc and a power-law temperature-density
relationship with $T_0 = 2 \times 10^4$ K and $\gamma = 1.6$.

Taking advantage of the larger volume of the $N$-body box, we create both 
a single mock reconstruction
spanning the entire $256^3$ box and 64 reconstructed subvolumes each with
dimensions
$32\,\hMpc{} \times 32\,\hMpc{} \times 256\,\hMpc{}$, 
which roughly match the CLAMATO survey geometry and volume.
The exact CLAMATO survey geometry 
($ 30\,\hMpc{} \times 24\,\hMpc{} \times 438\,\hMpc{}$) cannot be reproduced 
even with the 256 \hMpc{} simulation, but it provides at least a rough
comparison.

We generated skewers using $640^3$ grids of the Ly$\alpha$ absorption field.  
We followed exactly the same procedures to generate mock CLAMATO observations
from the $N$-body simulations as from the hydrodynamic simulations.

\vspace{50pt}

\section{Void Finding}
\label{sec:void_finding}

\subsection{Calibrating the void finder}
\label{sec:calib_vf}
To identify cosmic voids in the IGM map, we use the void-finding procedure described in 
\citet{stark:2015a}, which is analogous to the spherical overdensity techniques used for halo-finding in N-body simulations but applied to underdensities.  While this
method cannot fully capture the complex and anisotropic
shapes of voids, it is simple, easy to use, and easy to 
apply to both the density field and flux field.
While alternative finders \citep[i.e. watershed methods;][]{neyrinck:2008} are widely used in the literature,
the complexity of these void finders may lead to poor performance in the presence of noise
in the tomographic maps \citep[e.g.][]{stark:2015a}. Moreover, 
as this is the first attempt at void detection in a qualitatively new data set,
the spherical overdensity finder has an attractive simplicity.

To identify voids, we begin by finding all points
with $\delta_F$ greater than some threshold\footnote{Recall that $\delta_F$
has a negative sign convention with respect to overdensities.}, or density
lower than a separate threshold (``SO threshold'').  Spheres are grown
around all these points until the average $\delta_F$ (density) in the sphere
reaches a second threshold (``SO average'').
All spheres with $r \leq 2$ \hMpc{} are removed and overlapping voids are eliminated by only keeping the void with the largest radius.

The SO threshold and SO average chosen in this paper are motivated
by the values given in Table 1 of \citet{stark:2015a}.
However, these thresholds are inapplicable to CLAMATO because they neglect
continuum error in the mock sightlines and do not match the mean flux of the observations.  
Continuum error is particularly important at the high-transmission (high $\delta_F$) end \citep{lee+15}.
By combining Equation~\ref{eqn:deltaf} and Equation~\ref{eqn:cont_err} and Taylor-expanding in 
the small quantity $\delta_{\textrm{cont}}$, the change in $\delta_F$ due to continuum error is $\delta_{\textrm{cont}} F/\langle F \rangle$;
thus continuum error is more important at the high-flux end than the low-flux end.
Moreover, since continuum error is correlated
along a sightline, it will both create spurious voids and erase real voids.
Since continuum error increases the spread of $\delta_F^{\textrm{rec}}$
at the high-flux end, adding continuum error will lead to more points
with extreme values of $\delta_F^{\textrm{rec}}$ and thus increase the void fraction.

Following \citet{stark:2015a}, we begin by finding voids in the
real and redshift-space density fields.  We use the same
real space thresholds as \citet{stark:2015a}, with SO threshold
of $\rho = 0.2 \bar{\rho}$ and SO average of $\rho = 0.4 \bar{\rho}$.
The SO threshold is derived from the central density of a void
at shell-crossing in the spherical top-hat collapse model
\citep{vp+11}; the SO average is less well-motivated,
and was chosen by \citet{stark:2015a} to best create
visually-identified voids surrounded by edges (i.e.
the bucket profile).
The values of the SO threshold and SO average in the redshift-space
density are arbitrary; we use the same values as \citet{stark:2015a},
$\rho_{\textrm{red}} = 0.15 \bar{\rho}$ for the SO threshold and
$\rho_{\textrm{red}} = 0.3 \bar{\rho}$ for the SO average.
We expect the thresholds to be lower in redshift space than in real
space due to outflows from voids.


  \begin{deluxetable}{cccc}
  \tablecaption{\label{tab:volume_fraction} Volume fraction for different
  void thresholds in simulated catalogs}
  \tablehead{
Field & SO thresh & SO average & Vol. frac.
}
\startdata
    $\rho$ & 0.2$\bar{\rho}$ & 0.4$\bar{\rho}$ & 0.180 \\ 
    $\rho_{\textrm{red}}$ & 0.15$\bar{\rho}$  & 0.3$\bar{\rho}$  & 0.173 \\ 
    $\delta_{F}$ & 0.192  & 0.152 & 0.180 \\ 
        $\delta_{F}^{\textrm{rec}}$ & 0.220  & 0.175 & 0.180 \\ 
                $\delta_{F}^{\textrm{rec}}$ ($N$-body) & 0.220  & 0.175 & 0.182 \\ 
\hline
CLAMATO & 0.220  & 0.175 & 0.195 \\
        \enddata
        \tablecomments{Comparison of volume fraction of voids in data and simulations (100 \hMpc{} hydrodynamic box and 256 \hMpc{} $N$-body box).  All simulated fields are from the hydrodynamic box unless otherwise
        noted. The simulated fields include real and redshift-space density fields, the underlying flux $\delta_F$, and the reconstructed
        flux $\delta_F^{\textrm{rec}}$, with CLAMATO-like sightline spacing and
        realistic noise and continuum error.  Both $\delta_F$ and $\delta_F^{\textrm{rec}}$ are adjusted to the mean flux used in CLAMATO at $z = 2.3$.}
  \end{deluxetable}

We find similar volume fractions in the $N$-body and hydrodynamic simulations
for voids in the real-space and redshift-space density fields \citep[17-18\% in hydrodynamic
simulations in Table~\ref{tab:volume_fraction} compared to 15\% in $N$-body from Table 1 in ][]{stark:2015a}.
The small remaining discrepancies may arise from
the slightly different cosmologies of the two simulations
and the fact that the $N$-body simulations neglect baryonic effects.

We choose the SO thresholds in the underlying flux field $\delta_F$ and the mock CLAMATO
reconstruction $\delta_F^{\textrm{rec}}$ to match the void fraction in the redshift-space
density field. These thresholds are listed in Table~\ref{tab:volume_fraction}.  
We do not use the same thresholds for $\delta_F$ as \citet{stark:2015a}, since we rescale $\langle F \rangle$ to $\langle F(z = 2.3) \rangle$ from \citet{fg:2008}, changing
the range of $\delta_F$ and necessitating the use of a different threshold.  This allows us to apply
the same SO thresholds to both the observations and the two simulations.
Furthermore, unlike \citet{stark:2015a}, we do not use the same SO thresholds for $\delta_F$
and $\delta_F^{\textrm{rec}}$, since the presence of continuum error substantially
broadens the PDF of $\delta_F^{\textrm{rec}}$, yielding a 24\% void fraction in $\delta_F^{\textrm{rec}}$
versus 18\% void fraction in $\delta_F$ for the same SO thresholds. Due to the sensitivity
of the void fraction to both the mean flux and the continuum error, we emphasize that picking
appropriate thresholds requires realistic mock reconstructions.
As a result, these thresholds are only applicable to the data presented in this paper.

  \begin{figure}[t!]
\centering{\psfig{file=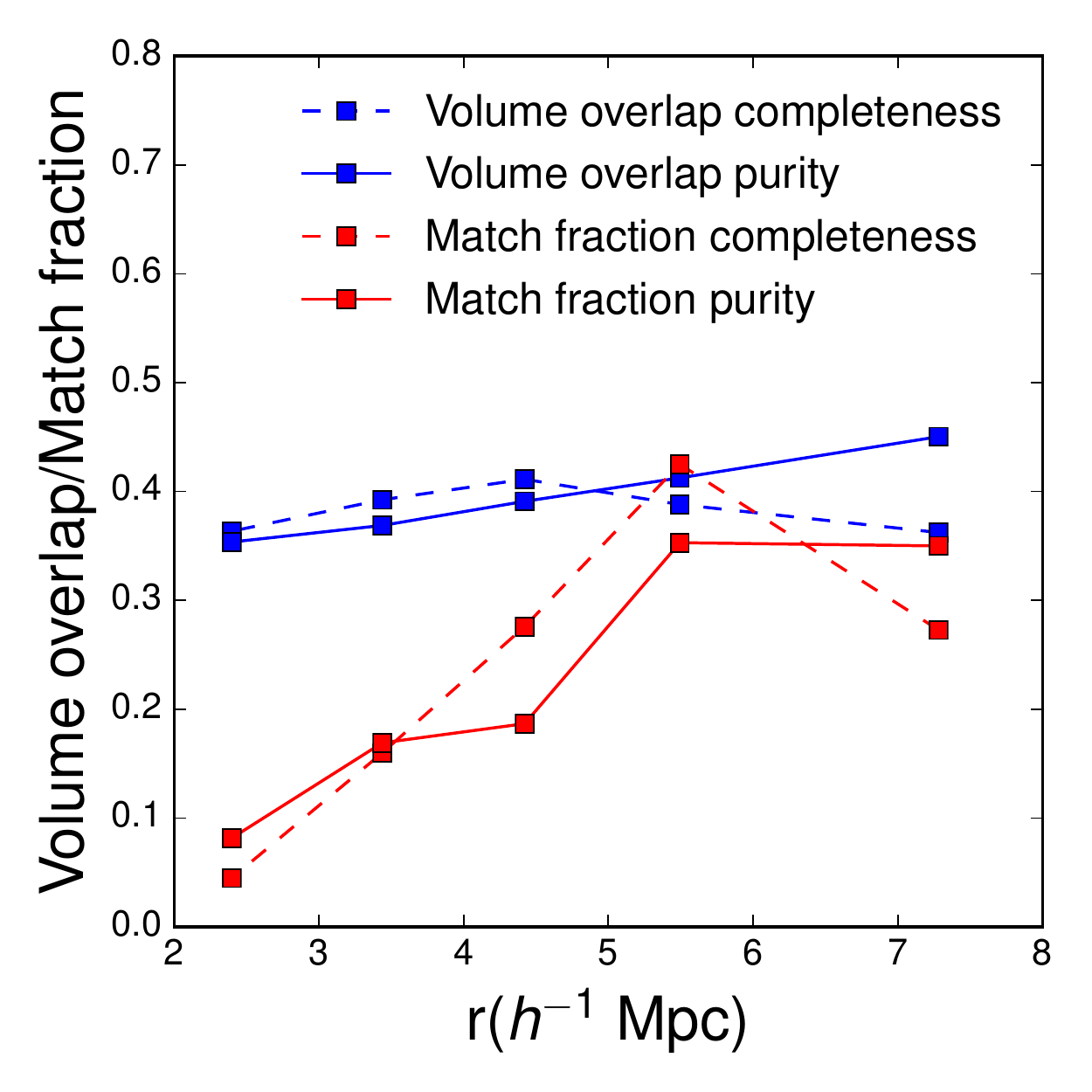,width=8 cm,clip=} }
\caption[]{\small Purity and completeness of the volume overlap fraction
and match fraction (fraction of voids with $\epsilon < 0.3$, see Equation~\ref{eqn:match_error})
as a function of void radius, measured between voids in the mock
CLAMATO-like reconstructions and the redshift-space density field
in the Nyx simulation.
}
\label{fig:volume_overlap_match_fraction}
\end{figure}

In the presence of continuum error, void recovery
is slightly poorer than reported in \citet{stark:2015a}.  
As in \citet{stark:2015a}, we characterize the fidelity of void recovery
using the volume overlap fraction and match error between redshift-space density field and mock-reconstruction voids.  The volume overlap fraction is defined as the fraction of the volume of voids in one catalog
that overlap voids in another catalog, while the match error is defined for each pair of voids A and B:
\begin{equation}
\epsilon = \frac{\sqrt{(r_A - r_B)^2 + |\vec{x}_A-\vec{x}_B|^2/3^2}}{r_A}
\label{eqn:match_error}
\end{equation}
For each void in catalog A, the match error is the minimum of the match error with all voids in catalog B.  Following
\citet{stark:2015a}, two voids are defined as well-matched if $\epsilon < 0.3$; thus the match fraction is the fraction
of all voids in a catalog with $\epsilon < 0.3$.

Depending on the comparison sample,
these quantities can describe either the purity or the completeness
of the void catalog: the completeness is characterized
by the (overlap or match) fraction of density voids that are also found in the reconstruction, while the purity is characterized
by the fraction of voids in the reconstruction that also exist
in the density field. 
We find that the completeness and purity drop 5 to 10 points compared to an identical mock observation
without continuum error.  Overall, we amend the conclusion
of \citet{stark:2015a} that 60\% of $r \geq 5$ \hMpc{}
voids are recovered by CLAMATO-like IGM tomography, instead finding the recovery of these large
voids to be closer to 40-45\%.

In Figure~\ref{fig:volume_overlap_match_fraction}, we plot the completeness and purity
of the volume overlap fraction and match fraction compared between voids in mock IGM tomography and the redshift-space density field in the Nyx simulation as a function of void 
radius. For large
voids, the completeness and purity of the match fraction and volume overlap fraction
range between 30 and 45\%
for $r \sim 6$ \hMpc{}.  For small voids, the match fraction drops rapidly 
to $\sim 5\%$ for $r \sim 2$ \hMpc{} while the volume overlap fraction drops more slowly, to
35\% for $r \sim 2$ \hMpc{}.  The same behavior was seen in \citet{stark:2015a}, and reflects
the fact that small voids may have poor centering and radius estimates due to tomographic noise artificially splitting or joining voids,
but the volume overlap fraction may nevertheless remain substantial.  We present
Figure~\ref{fig:volume_overlap_match_fraction} as a guide for using the void catalog
(Table~\ref{tab:all_voids}).  In Section~\ref{sec:void_prop}, we only use the high-quality $r \geq 5$ \hMpc{} sample
for studying void profiles, as this sample is less contaminated by noise in the tomographic reconstruction.

\subsection{Application to data}
\label{sec:app_to_data}
\newcolumntype{C}{>{\centering\arraybackslash}p{4em}}

\begin{figure*}
\includegraphics[width=\textwidth,clip=true,trim=500 20 500 120]{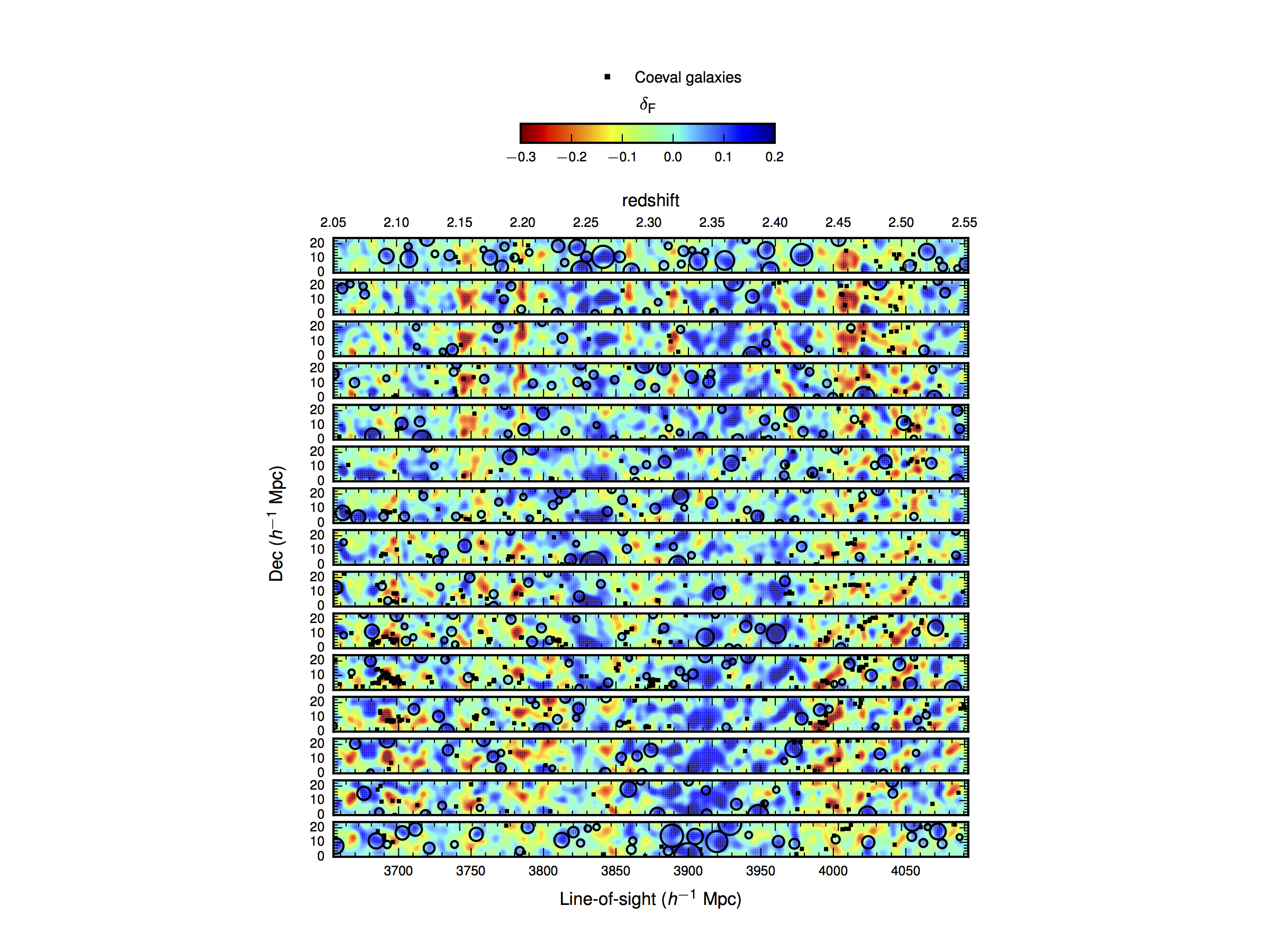}
\caption[]{\small Voids (circles) and spectroscopic galaxies (squares)
in the 2017 CLAMATO map.  Blue indicates regions of low absorption and thus low density
and high $\delta_F$, while red indicates regions of high absorption, high density, and low $\delta_F$. Each strip is a slice through the
RA direction, spaced by 2 \hMpc{} (strips are centered at 
RA = 1 \hMpc{}, 3 \hMpc{}, etc.).  RA increases from
the bottom strip to the top strip and declination increases from bottom to top on each strip.
In each strip, we plot voids between 0 and 2 \hMpc{},
2 and 4 \hMpc{}, etc.  Note that we only plot voids on the strip
where they are centered, although they may span
multiple strips.
}
\label{fig:voids_galaxies}
\end{figure*}

Applying the SO void finder to the 2017 CLAMATO IGM tomography map \citep{lee+17}, we identify 355 $r > 2$ \hMpc{} cosmic voids, including 48 higher-quality
$r \geq 5$ \hMpc{} voids which we use for studying the void profile (Section~\ref{sec:void_prop}).
These voids fill 19.5\% of the tomographic volume.  Table~\ref{tab:all_voids} presents the radii and positions of the voids
in both sky coordinates and tomographic map coordinates.
In Figure~\ref{fig:voids_galaxies},
we overplot the voids and positions of coeval spectroscopic galaxies
from MOSDEF, VUDS, zCOSMOS, and our own survey 
(see Section~\ref{sec:void_galaxy_cic} for descriptions
of these surveys).
The figure shows slices through the volume,
sampled every 2 \hMpc{} in the right ascension or longitudinal direction.  
While most voids span more than one slice in this plot,
for clarity we only show voids in the slice where their respective centers
are located.  Voids in Figure~\ref{fig:voids_galaxies}
appear largely devoid of galaxies, though a visual
evaluation of the galaxy distribution is difficult
owing to the very non-uniform selection function of the coeval 
galaxy spectroscopy.
A quantitative analysis of galaxies
within the tomography-identified voids is presented in Section~\ref{sec:void_galaxy_cic}.

\begin{figure*}[htb!]\centering
\includegraphics[width=\textwidth,clip=true, trim=20 20 10 0]{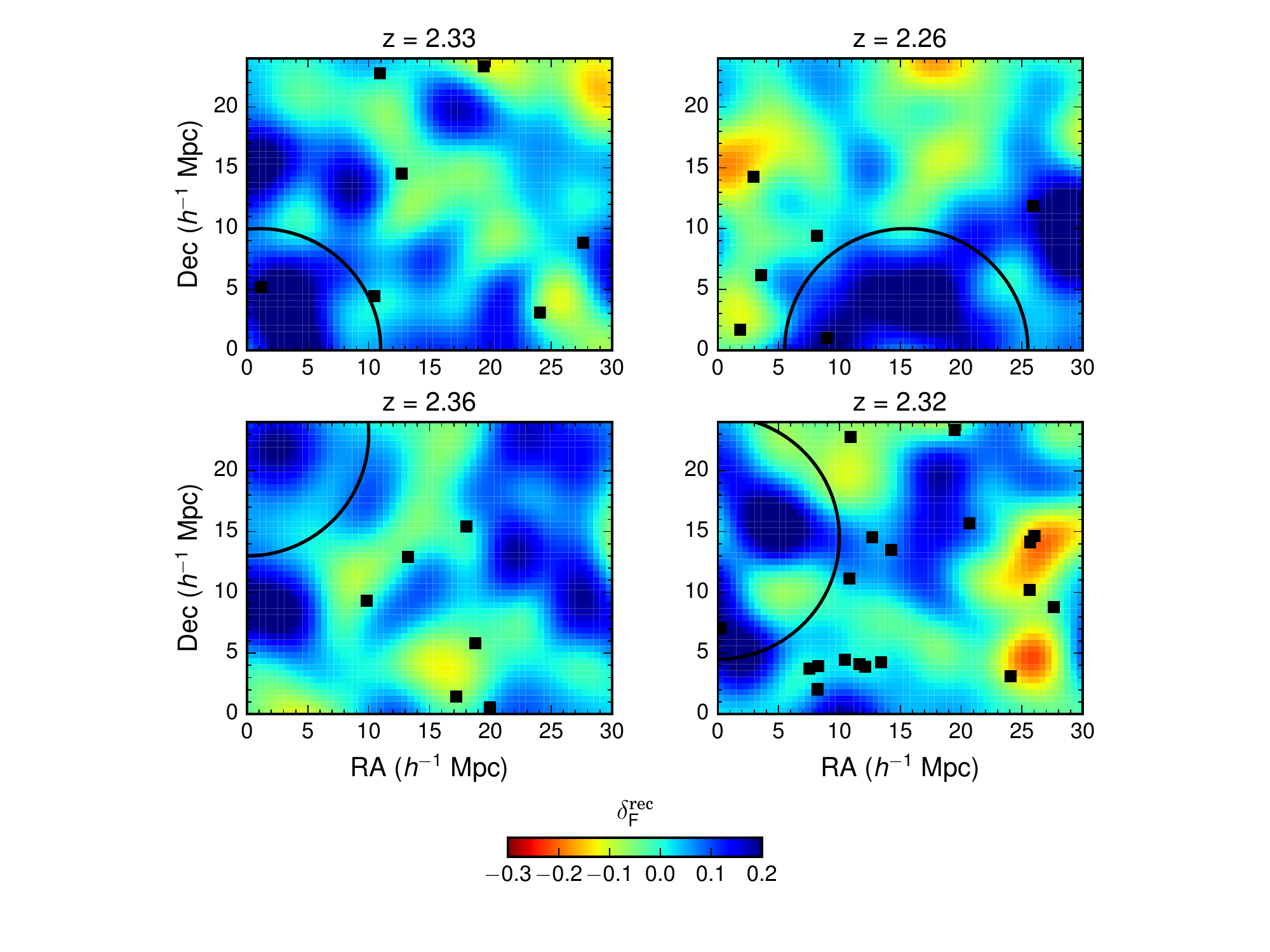}
\caption[]{\small Projections of $\delta_F^{\textrm{rec}}$ onto the line of sight for the four
largest voids in Table~\ref{tab:all_voids}.  In each panel, we plot the mean $\delta_F^{\textrm{rec}}$ averaged along a 20 \hMpc{}
length along the line of sight,
centered at the redshift of each void.  The black circle shows the void and black squares
are coeval galaxies within $\pm 10$ \hMpc{} of the void center.  
}
\label{fig:individual_voids}
\end{figure*}

Figure~\ref{fig:individual_voids} shows projections onto the plane of the sky for the four largest voids in our volume.  In each projection, $\delta_F^{\textrm{rec}}$ is averaged across
20 \hMpc{}
along
the line of sight (roughly the diameter of these voids).  We show all coeval
galaxies within this slice; therefore, galaxies with a different
redshift from the void center may appear to lie in a void in Figure~\ref{fig:individual_voids}
while actually lying outside the void boundaries.

We highlight a complex of several voids between
RA 0 and 10 \hMpc{}, declination 0 and 20 \hMpc{},
and $z = 2.32 - 2.37$.  While this structure is
broken into many voids by the spherical void
finder, it is likely that these voids are part of a single
structure spanning 10-20 \hMpc{}, including the largest
single void in the map, located at $(x,y,z) = (1,0, 244.5)$ \hMpc{} with radius 9.40 \hMpc{}.
As this void is located at the very bottom of the map, future observations extending the map will better probe
this structure.

\begin{figure}
\centering{\psfig{file=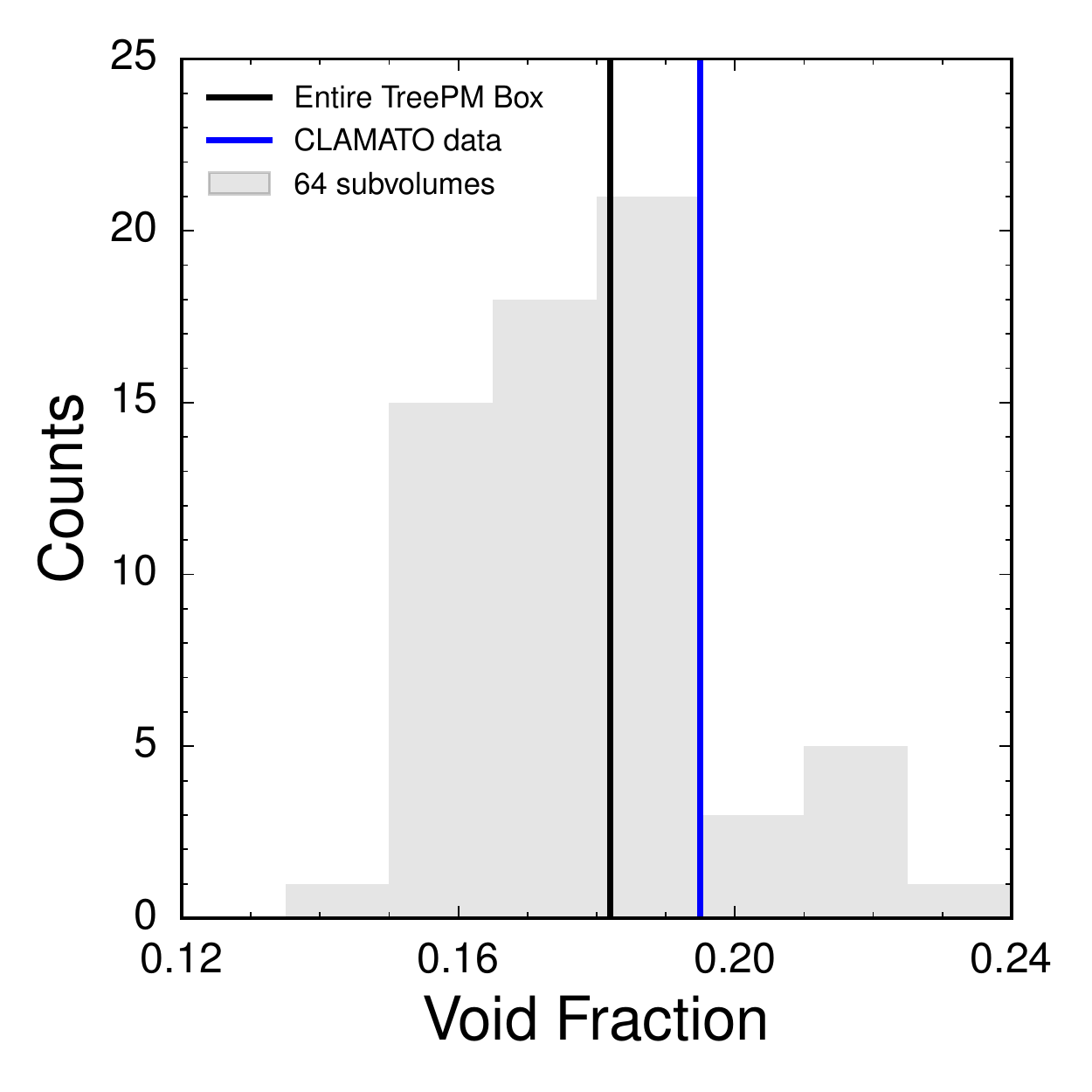,width=8 cm,clip=} }
\caption[]{\small Volume fraction of cosmic voids from 64
subvolumes each with dimensions $32 \times 32 \times 256$ \hMpc{} (light gray histogram), 
extracted from the $N$-body
256 \hMpc{} simulation box, compared to
void fraction from entire box (black) and in CLAMATO data (blue).
}
\label{fig:void_fraction_cosmic_variance}
\end{figure}

While the void fraction in CLAMATO (19.5\%) is slightly higher than the
void fraction in the mocks (18\%), this difference can be entirely explained
by sample variance.
To quantify the impact of sample variance on the void fraction,
we compute the void fraction in 64 subvolumes from our 256 \hMpc{} $N$-body simulation.
We find that the void fractions in the subvolumes range from 14.5\% to 22.8\%,
with a mean of 17.9\% and a standard deviation of 1.8\% (Figure~\ref{fig:void_fraction_cosmic_variance}). 
The small difference between the mean void fraction of this sample and the void fraction of the 
full $N$-body box (18.2\%) is attributable to the effects of an elongated geometry
on the $N$-body subvolumes, and suggests that further deviation from the mean void fraction of the subvolumes
due to the difference in survey geometry between CLAMATO and the $N$-body subvolumes is negligible.
The void fraction in the
CLAMATO map is thus $\sim1\sigma$ higher than the void fraction in the $N$-body
and hydrodynamic mocks.

In principle, matching the void fraction and void statistics
in the simulation requires matching Ly$\alpha$ statistics such as the flux PDF and the
flux power spectrum.  In practice, matching the flux PDF
especially is notoriously difficult, creating an additional
source of systematic error that may lead to disagreement
between void-finding in data and in simulations.  
Moreover, discrepancies between theory and data are especially
significant at the high-transmission end of the PDF, $F > 0.8$,
where the voids lie \citep{bolton+16}. The high-transmission
end of the flux PDF is particularly sensitive to the slope
of the temperature-density relationship $\gamma$ \citep{white+10}.
Early measurements of the flux PDF suggested that $\gamma \lesssim 1$ \citep{bolton+08}, in contrast to $\gamma \sim 1.6$ used
in simulations here, though \citet{lee12} pointed out that
the effects of continuum error can be degenerate with changing $\gamma$.  Later measurements of the flux PDF from BOSS with better controlled
continuum fitting found $\gamma \sim 1.6$ \citep{lee+15},
though \citet{rorai+17} claim that even with improved continuum
fitting, high-resolution quasar spectra still favor $\gamma \lesssim 1$,
especially in underdense regions.  
Overall, \citet{lee+15} showed that careful modeling of noise
and systematic errors are critical for interpreting the flux
PDF of low-resolution, noisy data such as BOSS or CLAMATO,
with spectral resolution, pixel noise and continuum error
playing a particularly prominent role.  They also find
that additional discrepancies remained at high flux, which 
they solved by varying $\langle F \rangle$.
Therefore, we carefully model pixel noise, continuum error,
and Gaussian smoothing from the LRIS spectrograph.
While we believe our current mocks are sufficiently realistic for an initial
void detection and characterization, more careful mocks will be required for future cosmological analyses of 
IGM cosmic voids.

\begin{deluxetable*}{CCCcccc}[t]
\tablecolumns{7}
\tablecaption{\label{tab:all_voids} Voids in CLAMATO 2017 Map}
\tablehead{
\multicolumn{3}{c}{Tomographic map position (\hMpc{})} & Void Radius & \multicolumn{3}{c}{Sky position} \\
$x$ & $y$ & $z$ & (\hMpc) & $\alpha$ (J2000) & $\delta$ (J2000) & redshift }
\startdata
1.0 & 0.0 & 244.5 & 9.40 & 149.96480 & 2.15000 & 2.33 \\
15.5 & 0.0 & 179.5 & 9.10 & 150.17943 & 2.15000 & 2.26 \\
0.0 & 23.0 & 273.5 & 7.90 & 149.95000 & 2.49016 & 2.36 \\
0.0 & 14.5 & 233.5 & 7.70 & 149.95000 & 2.36445 & 2.32 \\
29.5 & 11.0 & 186.0 & 7.65 & 150.38665 & 2.31268 & 2.26 \\
23.0 & 0.0 & 366.0 & 7.45 & 150.29044 & 2.15000 & 2.47 \\
29.5 & 12.5 & 323.0 & 7.40 & 150.38665 & 2.33487 & 2.42 \\
0.0 & 10.5 & 264.5 & 7.25 & 149.95000 & 2.30529 & 2.35 \\
29.5 & 1.0 & 171.0 & 7.00 & 150.38665 & 2.16479 & 2.25 \\
3.5 & 0.0 & 293.0 & 6.95 & 150.00181 & 2.15000 & 2.39 \\
\enddata
 \tablecomments{Table 2 is published in its entirety in the machine-readable format.
      A portion is shown here for guidance regarding its form and content.}
\end{deluxetable*}

\vspace{20pt}

\section{Void-Galaxy Counts in Cells}
\label{sec:void_galaxy_cic}

\begin{figure*}[t]
\begin{tabular}{cc}
\centering{\psfig{file=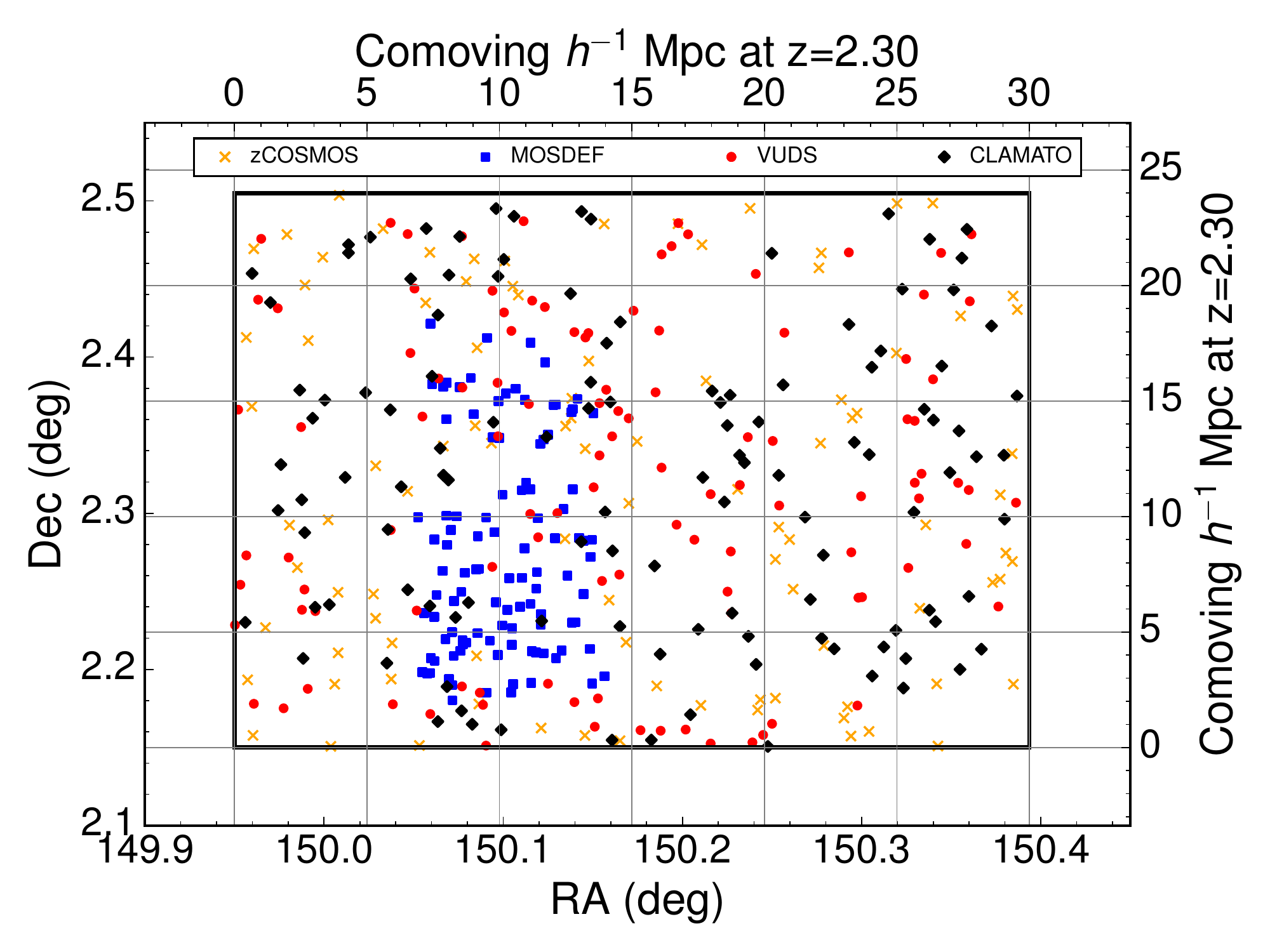,width=9 cm,clip=} }
& \centering{\psfig{file=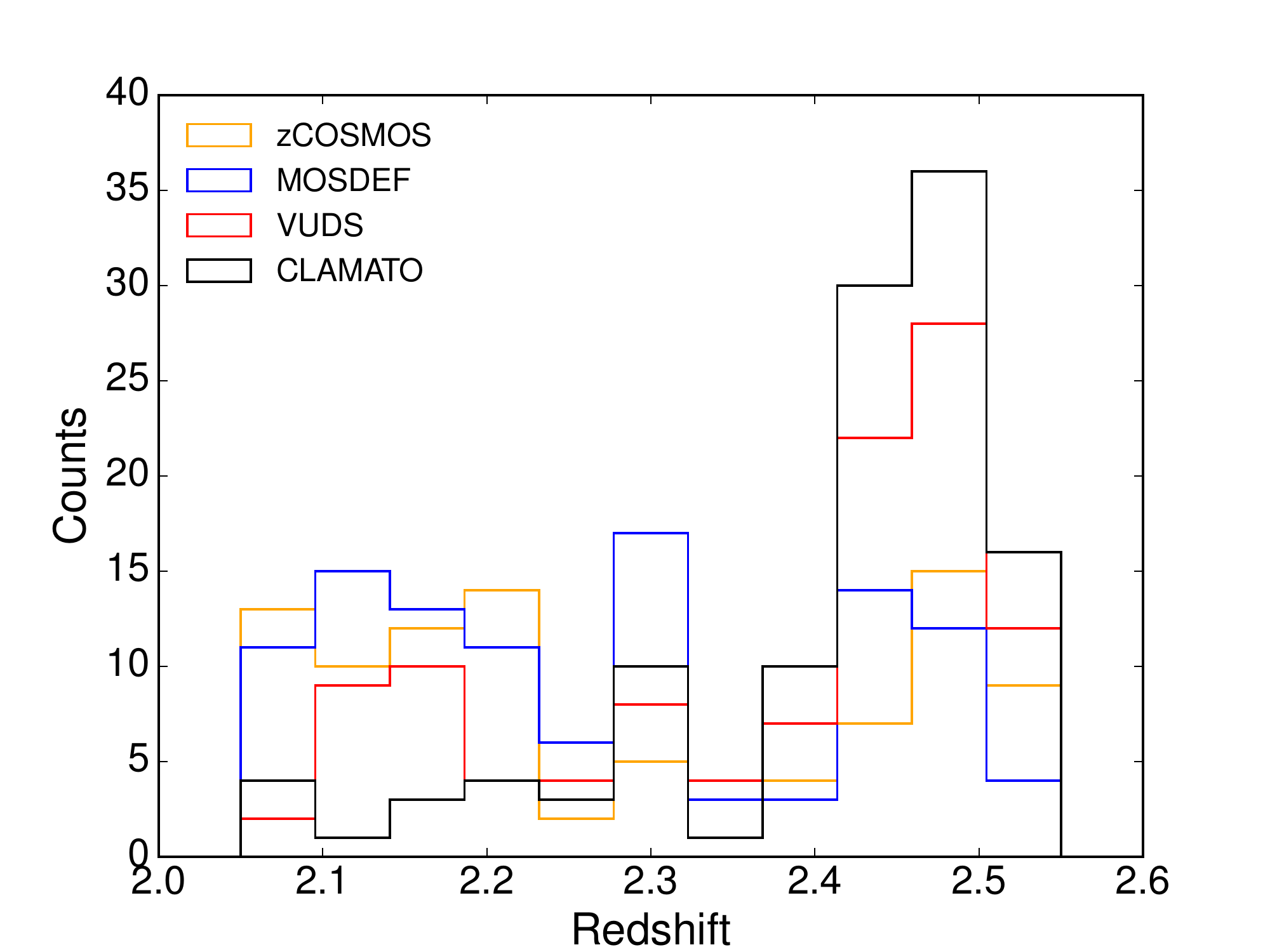,width=9 cm,clip=} }
\end{tabular}
\caption[]{\small \textit{Left: } Positions of galaxies in the COSMOS field with known spectroscopic redshifts that are coeval with the $2.05<z<2.55$ CLAMATO map.
The black box indicates the footprint of the CLAMATO map.
\textit{Right: } Redshift distribution of coeval galaxies.
}
\label{fig:coeval_gals}
\end{figure*}

The cosmic voids in the CLAMATO volume are by far
the most distant sample of cosmic voids known at the present time.
In comparison with the most distant $z\sim 1$ voids previously 
detected in galaxy redshift
surveys \citep[e.g.,][]{conroy+05,mich+14}, our voids at $z\sim 2.3$ are $\sim 1.7\times$
further in terms of comoving distance.
Moreover, since CLAMATO achieves $\gtrsim3$ times
better density field resolution than existing or upcoming galaxy surveys at $z \sim 2$,
it represents the best method for detecting high-$z$ voids for the immediate future
\citep[although all-sky interferometric 21cm surveys may be able to detect voids at $z \sim 1-2$; ][]{white_padmanabhan:2017}.

  \begin{deluxetable*}{ cccccc }
  \tablecaption{\label{tab:void_galaxy_cic} Significances of galaxy underdensities
  in voids}
  \tablehead{
Galaxy survey & $N_{gal}$ & Galaxies in voids & Galaxies in randoms (mean) & Galaxies in randoms ($\sigma$) & Significance
}
\startdata
 VUDS & 110 & 13 & 20.36 & 4.29 & 0.0491 \\
MOSDEF & 109 & 6 & 18.49 & 5.45 & 0.0047 \\
CLAMATO & 118 & 10 & 22.07 & 4.67 & 0.0033 \\
zCOSMOS & 95 & 8 & 18.63 & 4.12 & 0.0035 \\
        \enddata
        \tablecomments{Significance of galaxy underdensities in 4 coeval galaxy
        surveys.  CLAMATO uses the galaxies spectroscopically confirmed by
        our data that lie within the map volume.}
  \end{deluxetable*}

\begin{figure*}
\centering 
\includegraphics[width=\textwidth]{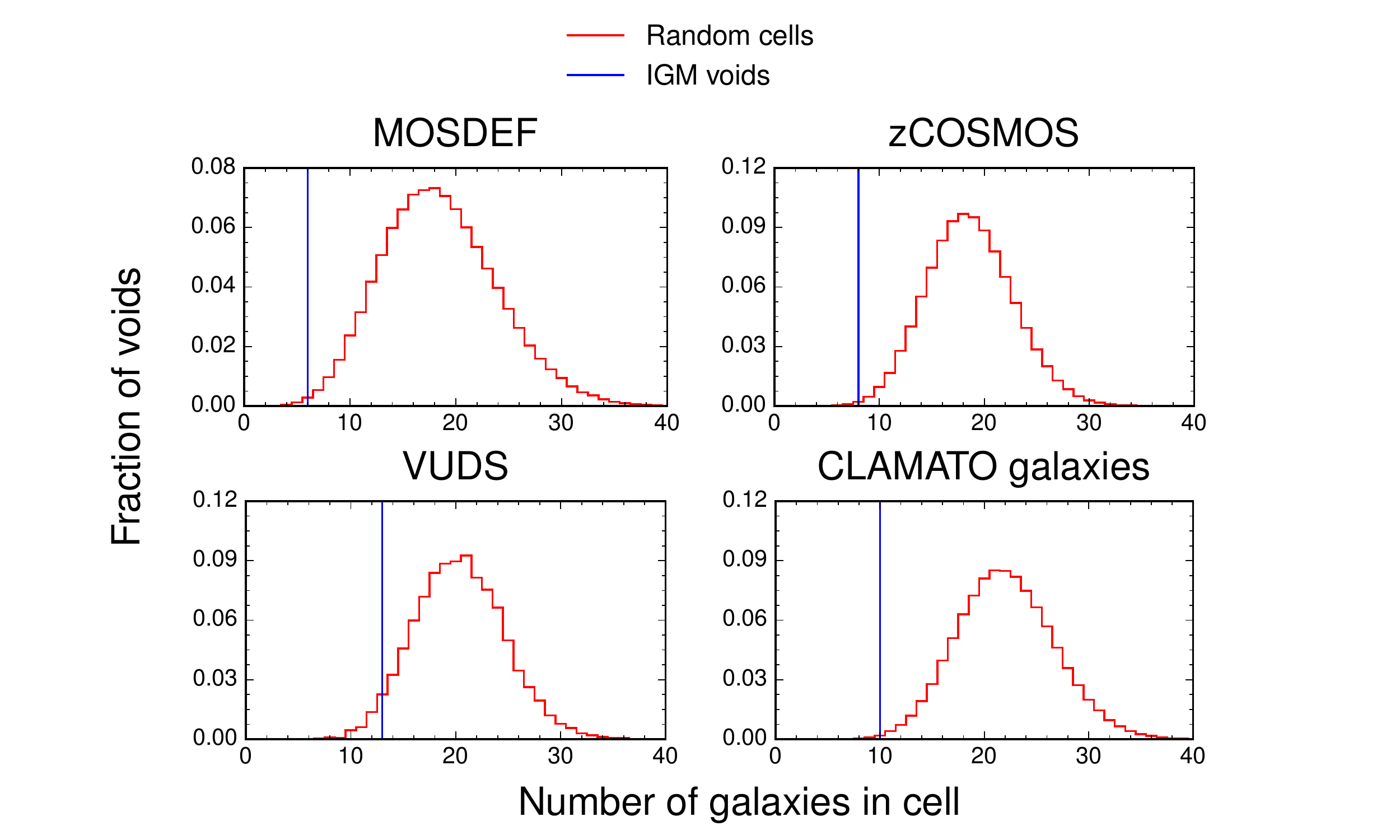}
\caption[]{\small Distribution of number of galaxies in random cells (red), compared to number
of galaxies in IGM voids (blue line), for 4 galaxy surveys.  The $p$ value is the fraction
of the red histogram to the left of the blue line.
}
\label{fig:n_random_n_voids}
\end{figure*}

We validate the void-finding technique by counting
coeval spectroscopic galaxies within the 
tomography-identified voids and comparing these counts
to the number of galaxies within random cells
with the same radius distribution and volume fraction.
Exploiting the rich set of spectroscopic data that already exists within the COSMOS
field, we use 110, 109, 118, and 95 galaxies from the VUDS \citep{le-fevre:2015}, 
MOSDEF\footnote{We use their 2016 August data release; \url{http://mosdef.astro.berkeley.edu/for-scientists/data-releases/}.}
\citep{kriek_mosdef}, 
CLAMATO, and zCOSMOS-Deep \citep{liil+07} surveys respectively, which directly overlap with the CLAMATO 
map volume at $2.05 < z < 2.55$. 
By CLAMATO, we mean galaxies that were spectroscopically confirmed by CLAMATO to lie inside the
map volume, e.g. sightlines for the lower redshift part of the map or galaxies
with redshifts too low to be viable sightlines.

In Figure~\ref{fig:coeval_gals}, we show the redshift distribution
of these coeval galaxies and their spatial coverage compared to the
CLAMATO area.  These surveys differ in their redshift accuracy:
the NIR-based redshifts from MOSDEF are most accurate \citep[][$\sigma_v \sim 60$
km s$^{-1}$, corresponding to $\sigma_{los} \sim 0.7$ \hMpc{}]{steid+10},
followed by the optical redshifts from VUDS, CLAMATO, and zCOSMOS
\citep[][$\sigma_v \sim 300$ km s$^{-1}$]{steid+10,kriek_mosdef}.
For this analysis, we do not include galaxies from two overlapping 
spectroscopic surveys, 3DHST and ZFIRE.
The grism redshifts from 3DHST have redshift uncertainties of 
$\sigma_v \gtrsim 500$ km s$^{-1}$
\citep{kriek_mosdef,momcheva:2016} which are comparable to the typical sizes
of our voids of a few cMpc. The ZFIRE survey \citep{nanayakkara:2016}
 specifically targeted at the $z \sim 2.1$ protocluster \citep{spitler+12} and is therefore a poor choice
for void validation because the galaxies will not lie in an average environment.

Galaxy positions are converted to $x,y,z$ coordinates with the
origin at $z = 2.05$, right ascension 149.95$^{\circ}$ and declination 2.15$^{\circ}$
using the transverse comoving distance evaluated at $z = 2.3$.
We convert galaxy redshift $z_{gal}$ to coordinate position $z$
using
\begin{equation}
z = (z_{gal} - 2.05) \left. \frac{d\chi}{dz} \right\rvert_{z=2.3}
\label{eqn:convert_z}
\end{equation}
Therefore, the conversion between ($\alpha$, $\delta$, $z$)
and map coordinates ($x,y,z)$ is identical for coeval galaxies
and Ly$\alpha$ forest pixels.

We emphasize that this comparison is simply a 
validation of the cosmic void sample, and that the void-finding 
on the tomographic reconstruction is entirely self-sufficient. 
Conversely, the spectroscopic redshift galaxy samples within the
field are too sparse and incomplete\footnote{We find $n_g \sim 1.1 \times 10^{-3} h^{3}\,\mathrm{Mpc}^{-3}$ for VUDS, CLAMATO, and zCOSMOS redshifts combined, 
compared to $n_g \sim 5 \times 10^{-3} h^{3}\,\mathrm{Mpc}^{-3}$ in the VIPERS survey which detected $z\sim 1$ voids.} to define cosmic voids, but 
should be sufficient to falsify a spurious detection of cosmic voids. 

To compare the abundance of galaxies in voids with
a control sample,
we create many realizations
of random catalogs with the same radius function
as the void catalog.
Many of the largest CLAMATO voids are preferentially
located near the edge of the CLAMATO volume.
Therefore, in order to reproduce the correct
volume fraction in the random catalogs,
we require each random cell to have the same
distance from the boundary
as the corresponding void with the same radius.

In detail, for each void in the catalog,
we create a random cell with the same radius.
If the void is located in a ``corner'' of the volume
(i.e. the distance between both its $x$ and $y$ positions and the box edge is smaller than the void radius), we assign
the random's $xy$ position by rotating the void's $xy$ position
about the origin by either $0^{\circ}$, $90^{\circ}$,
$180^{\circ}$, or $270^{\circ}$.  We then randomly
assign the $z$ position.  For voids not located
in a corner, we randomly assign
the position along the faces of a rectangular prism
with distance to the CLAMATO volume edge
equal to the minimum distance between the
void center and the box edge.  Just like the voids, the random
cells are required to be non-overlapping.
We find that the random cells fill 18.9\% of the CLAMATO
volume (on average), compared to 19.5\% of the CLAMATO volume
filled by voids.
As a sanity check that the random cells are indeed unbiased regions, 
we also find in the random cells
an average absorption of 
$\langle \delta_F^{\textrm{rec}} \rangle = -6.57 \times 10^{-3} \pm 5.1\times 10^{-3}$ (1$\sigma$ standard deviation) 
compared
to $\langle \delta_F^{\textrm{rec}} \rangle = -7.23 \times 10^{-3}$ for the entire map.
In other words, they are both consistent with zero as would be expected
by definition (Equation~\ref{eqn:deltaf}).

For MOSDEF, we use separate random catalogs covering the smaller area
probed by this survey (Figure~\ref{fig:coeval_gals}) 
rather than the entire CLAMATO volume.  This allows the random catalogs to accurately reproduce
the void fraction within the MOSDEF survey region. We use an area that
extends 3 \hMpc{} beyond the approximate MOSDEF footprint: in this case,
$150.001\degr > \mathrm{R.A.} > 150.203\degr$ and $2.150\degr > \mathrm{Dec} > 2.444\degr$.  We include voids that are
slightly outside the MOSDEF footprint because these voids may
still overlap with MOSDEF galaxies; we choose a 3 \hMpc{} buffer
because the average void size is about 3 \hMpc{}.
The random cells fill 12.5\% of the
MOSDEF region volume, compared to 12.6\% void fraction in this region,
with average
$\delta_F^{\textrm{rec}}$ $-0.0113 \pm 0.0091$.
The smaller void fraction may be due to the fact that the MOSDEF region is slightly
overdense, with $\langle \delta_F^{\textrm{rec}} \rangle = -0.0121 \pm 0.0002$ (standard
error of the mean)
compared to $\langle \delta_F^{\textrm{rec}} \rangle = -0.0073 \pm 0.0001$ in the entire map.

The significance of the galaxy underdensity in tomographic voids
is the probability that the number of galaxies in random cells
is less than or equal to the number of galaxies in voids.  We calculate this probability by counting the number
of realizations of the random catalog with fewer galaxies
in the randoms than in the tomographic voids, giving a
$p$ value for each galaxy survey.  Assuming that the constraints
from the different galaxy surveys are independent,
the combined constraint is simply the product of
the $p$ values for the individual surveys.
The distribution of galaxy counts in random cells is neither Gaussian
nor Poissonian, particularly as it approaches zero galaxies where
the data lies; therefore, calculating $p$ values by direct simulation
is essential and we emphasize that the conversion to $\sigma$
is purely for illustrative purposes.
The error on $p$ values computed this way are given by
$\sqrt{p(1-p)/N}$.
In order to achieve $<10\%$ errors on $p$ values, we use
10,000 realizations of the random catalog for VUDS, and 300,000 realizations for CLAMATO,
MOSDEF, and zCOSMOS.
 

We report significances in Table~\ref{tab:void_galaxy_cic}
and compare the number of galaxies in voids to the number of galaxies
in random cells in Figure~\ref{fig:n_random_n_voids}.
Assuming that the galaxy surveys are independent, we find a combined
$p$ value of $3 \times 10^{-9}$, equivalent to a $5.95\sigma$ detection
of galaxy underdensities in the tomography-identified voids.

The significance of the galaxy underdensity in the tomographically-identified
voids is similar for all four surveys, although modestly lower for VUDS.
These galaxies are the faintest of the surveys
used ($ \langle r \rangle = 24.9$, compared to $\langle r \rangle = 24.1$ for CLAMATO and zCOSMOS and $\langle r \rangle = 24.8$
for the primarily quiescent MOSDEF sample), and are thus likely have lower bias, causing them to cluster
towards voids \citep{conroy+05}.

\section{Void properties}
\label{sec:void_prop}

\subsection{Void radius function}
\label{sec:vrf}

\begin{figure*}
\centering{\psfig{file=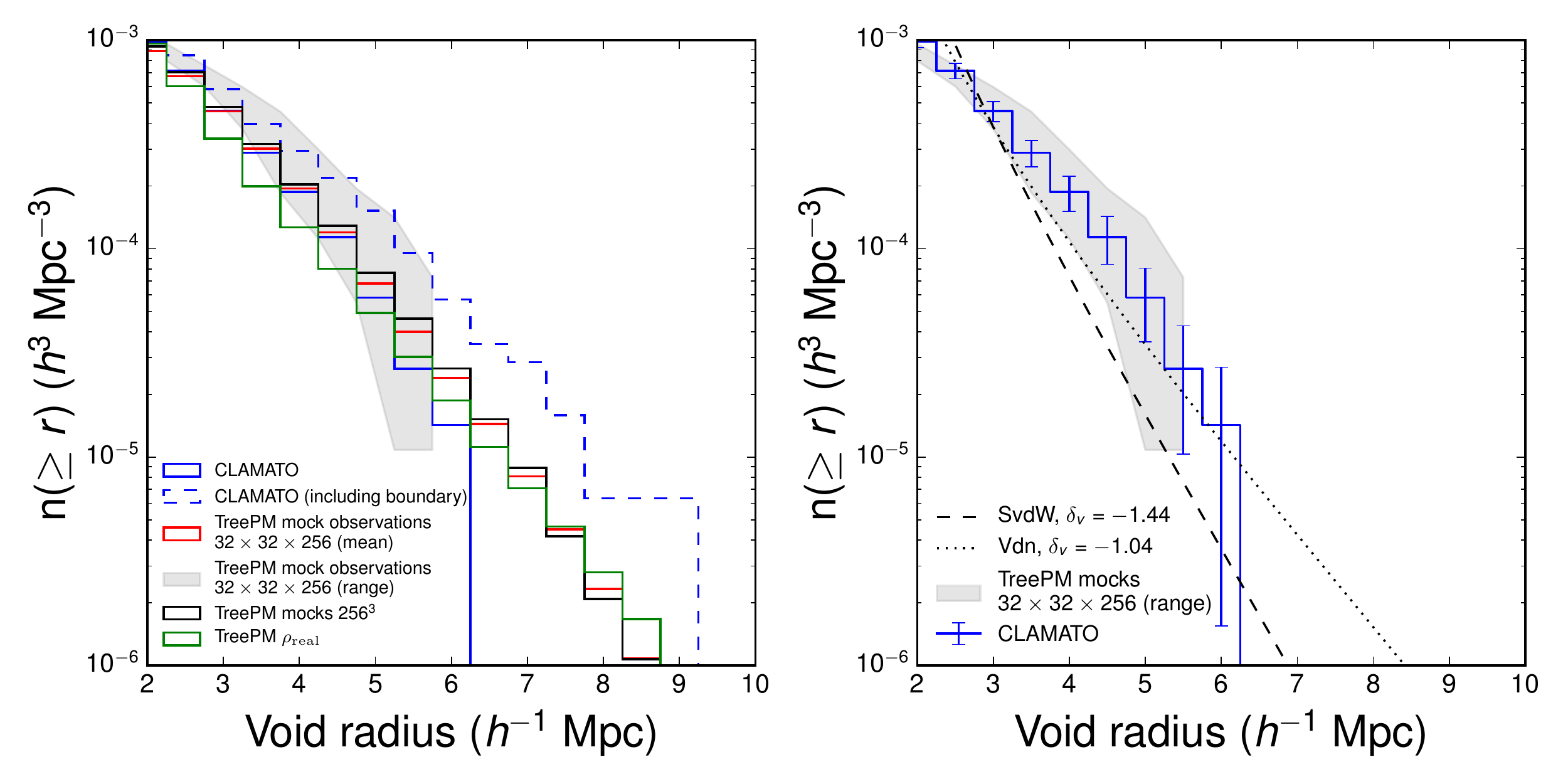,width=18 cm,clip=} }
\caption[]{\small \textit{Left:} Comparison of void radius function in CLAMATO to void radius function
from the $N$-body real-space density field; the mean and range of the void radius function from mock observations
constructed from 64 $(32 \times 32 \times 256)$
\hMpccube{} subvolumes of the N-body box; and the void radius function from a mock observation constructed from the full $256^3$
\hMpccube{} box.  In all cases we exclude voids with distance to the boundary smaller than the void radius, except
for the blue dashed line, which gives the abundance of all CLAMATO voids and thus shows the impact of edge effects on the CLAMATO void abundance.
In all cases, we have centered each histogram bin
over the corresponding void radius: i.e. the bin centered at
3 \hMpc{} gives the number of voids with radius greater than or
equal to 3 \hMpc{}.
\textit{Right:} Comparison of the CLAMATO void radius function to excursion set models (black lines),
with the range of the 64 $N$-body subvolumes overplotted to give a sense of the error on the measured
void radius function.  Error bars on the data are Poisson
error bars on the counts in each bin, divided by the effective
volume of that bin.
}
\label{fig:void_radius}
\end{figure*}

\begin{figure}
\centering 
\psfig{file=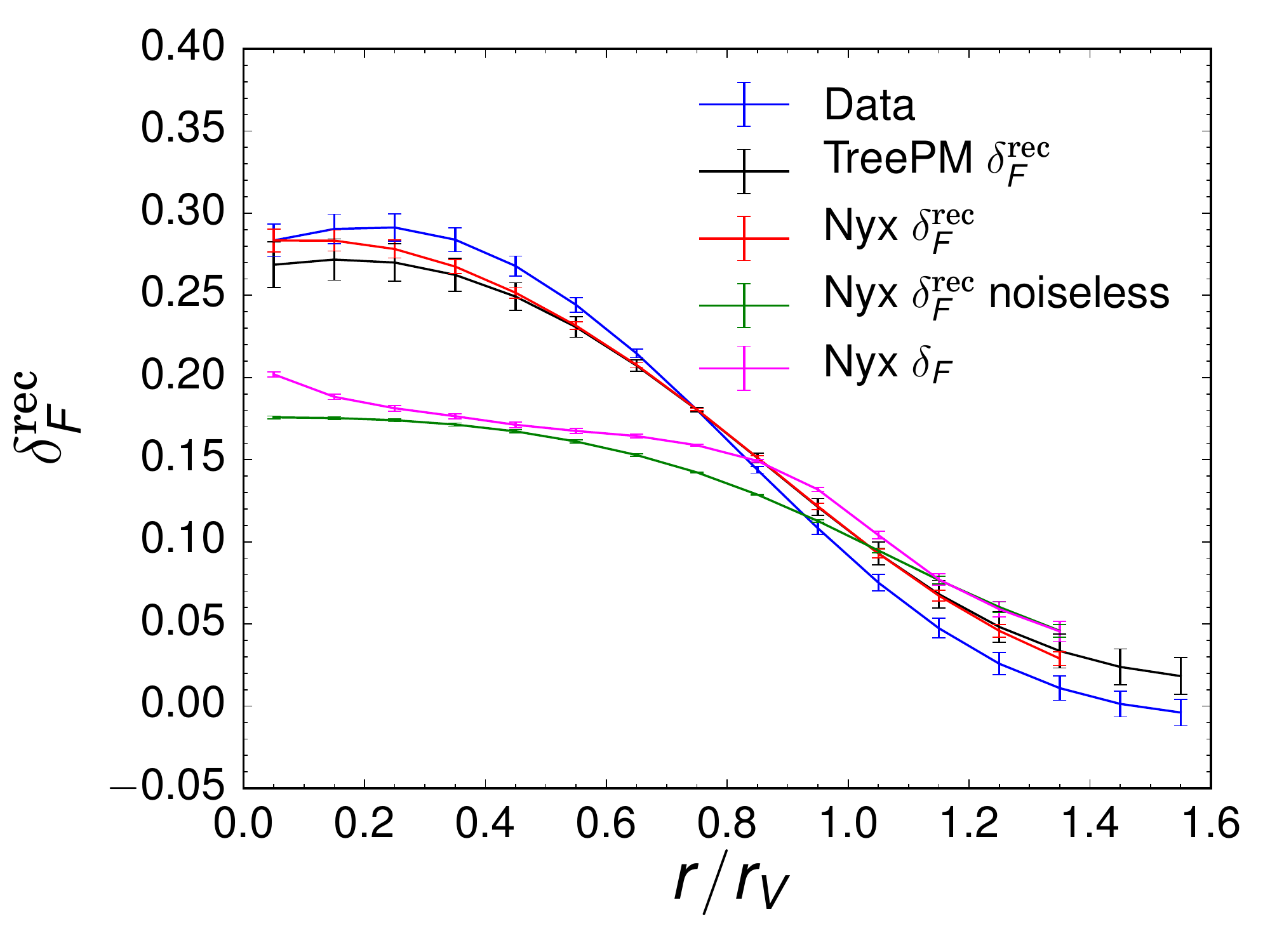,width=8 cm,clip=} 
\caption[]{\small Radially-averaged void profiles in data (blue), mock observations (black for $N$-body box and red for hydrodynamic box), noiseless mock observations (green) and underlying $\delta_F$ (magenta),
stacked in units of the void radius $r_V$, for all voids
with $r \geq 5$ \hMpc{}. Errorbars for the Nyx profile are estimated using 1000 realizations
of the void catalog generated via bootstrap resampling, while errorbars for the TreePM and data profiles are generated
from the standard deviation over the 64 subvolumes of the TreePM box.
}
\label{fig:void_profile1d}
\end{figure}

We compare the void radius function in CLAMATO to the void radius function
in mock observations and in the real-space density field (Figure~\ref{fig:void_radius}).
Due to edge effects, voids are significantly more likely to be found near
the map boundaries of both the CLAMATO data and the 64
subvolumes of the $N$-body box with roughly CLAMATO-like geometry.  
As a result, we omit voids found within one void radius of the box edge.
To compute the void radius function, we weight
each void by the effective volume over which it could have been observed: for a void of radius $r$, this volume is $(30-2r)(24-2r)438$ $h^{-3}$ Mpc$^{3}$.  Omitting voids near the box edge
leads to substantially
better agreement with the void radius function in both the density field
and the full-volume reconstruction.
We also overplot the range of void radius functions found in the 64 subvolumes
to give an estimate of the impact of sample variance on this measurement;
we do not plot the range for large voids where the Poisson errors become large
due to the relatively small volume of both the CLAMATO and simulated survey volumes.

We compare the measured void radius function to predictions from excursion set theory \citep{sheth_vdw:2004,jennings:2013}.  The excursion set model associates voids
with spherical regions that have just undergone shell crossing and have thus attained
an average density of $0.2 \bar{\rho}$.  The evolution of voids is modeled as a random
walk with two barriers, a lower barrier at $\delta_v = -2.71$, the linear underdensity
of shell-crossed voids, and an upper barrier, $\delta_c$, modeling voids squeezed out of existence
by surrounding overdensities, ranging between 1.06 and 1.69.  We fit two excursion set models
to the data, the number-density-preserving model of \citet{sheth_vdw:2004} (SvdW)
and the volume-preserving model of \citet{jennings:2013} (Vdn).  $\delta_v = -2.71$ provides
a poor fit in both cases, so we allow the void threshold to vary as a free parameter,
finding $\delta_v = -1.44$ ($-1.04$) for the SvdW (Vdn) models.
We use $\chi^2$ minimization to determine the best-fit $\delta_v$,
with error bars given by the Poisson errors on the number of voids in
each bin divided by the effective volume of that bin, i.e.
$(30-2r)(24-2r)438$ $h^{-3}$ Mpc$^{3}$ for a bin at radius $r$.  Owing to the large range
in $n(\geq r)$, we minimize $\chi^2$ in log-space rather than linear
space.  We find that neither model can adequately explain
the void radius function at small $r$ ($< 3$ \hMpc{}), where the
error bars are substantially smaller than at large $r$.  As a result,
the best-fit curves for both models are ``tilted'' relative to the
data at $r \geq 3$ \hMpc{} due to the smaller errorbars and thus
larger impact of the points at small $r$.
While neither model can fit the void radius function at small
radii, the Vdn model adequately fits the data at large $r$, and provides
a notably better fit than the SvdW model.

We expect a higher value of $\delta_v$ than -2.71 for the void radius function
in our work because we use a higher mean overdensity of voids
($\bar{\rho} = 0.4$); indeed, our results are similar to the results
of \citet{jennings:2013}, who found $\delta_v = -1.24$ for $\bar{\rho} = 0.4$.
Our results also lie in the same general range as previous results,
which find $\delta_v$ between -0.2 and -1.0 \citep{sutter:2014b,pis+15,nad+hotch:2015}.
However, \citet{jennings:2013}, working
between $z = 0$ and 1, recommend models with considerably
smaller void abundance than found here ($1/5$ the abundance of the
\citet{sheth_vdw:2004} prediction with $\delta_v = -1.24$, about 5
times lower than our data).


\vspace{10pt}

\subsection{Radial void profile}
\label{sec:profile}


We plot radially-averaged void profiles in Figure~\ref{fig:void_profile1d}
for all voids with $r \geq 5$ \hMpc{}, normalizing each void
to its void radius and stacking in units of the void radius $r/r_V$.
There is good agreement between void profiles in data and mock observations, with $\chi^2 = 22.1$ over 16 radial bins
between the void profile in CLAMATO and the void profile in mock observations from the $N$-body simulations. 
Since each bin is $0.1 r_v \sim 0.5$
\hMpc{}, much smaller than $\langle d_{\perp} \rangle = 2.5$ \hMpc{},
the void profile is highly correlated between neighboring bins, so we
cannot assume a diagonal covariance matrix when computing $\chi^2$ (i.e. the $\chi^2$
quoted above uses the full covariance matrix and is much lower than if this covariance
matrix were diagonal).  We compute the covariance matrix
using the 64 subvolumes of the $N$-body box and scale down the covariance by $0.8$, the volume ratio between the $N$-body subvolumes
and the CLAMATO volume.
We also use the unbiased estimator of \citet{hart+06} for the inverse covariance matrix,
for the case where the mean is estimated from the data (their Equation 17).
The strong agreement between the radial void profile in mock observations
and data suggests that approximations in the map-making process
(e.g. the distance-redshift and angle-redshift conversion discussed in Section~\ref{sec:data}) make only a minor impact on the void profile.

The void profile in mock observations
traces the void profile in the underlying Ly$\alpha$ flux field, $\delta_F$, well for $r > r_V$ but deviates
badly inside the void.  This deviation is due almost entirely to noise
in the spectra, with the profiles in noiseless reconstructions resembling
the $\delta_F$ profiles much more closely.  Unfortunately, the deviation
between void profiles in $\delta_F^{\textrm{rec}}$ and $\delta_F$
means that void profiles in the reconstruction do not trace void
profiles in matter, and thus we do not try to fit a functional
form to the void profile \citep[e.g.][]{cecc+06,ham+14,white_padmanabhan:2017} as it could
not be compared with low-redshift results.

Qualitatively, the void profile in the data is missing the ``compensation wall'' that is present in some low redshift void profiles, particularly voids
with $r < 20$ \hMpc{} like those discussed here \citep{ham+14}.
It is unclear whether the absence of a compensation wall is indicative
of physical differences between high and low redshift voids, or is merely
an artifact of our void finder and void sample. For instance, while our voids
are small at $z \sim 2.3$, they will become much
bigger by $z \sim 0$: \citet{sheth_vdw:2004} find 
$r_v \propto (1+z)^{-2/(3+n)}$, where $n \sim -1.5$ is the slope of the power spectrum
on scales of the void size.  Therefore  5 \hMpc{} voids at $z = 2.3$ correspond to 25
\hMpc{} voids at $z = 0$, which generally have a very weak or absent
compensation wall \citep{cai:2015,ham+16}. On the other hand,
\citet{white_padmanabhan:2017} suggest that spherical overdensity finders
may not find compensation walls, while \citet{cai:2016} argue that compensation
walls are only present in voids found in overdense environments.

We also study the impact of redshift-space
distortions on $z\sim2$ voids.  Redshift-space distortions modify the void profile along the line of sight
and are often measured using the quadrupole of a correlation
function or void profile.
Numerical simulations find that for $r \gtrsim r_V$
in uncompensated voids, iso-density contours
are flattened along the line of sight in the same
sense as the \citet{kaiser:1987} effect for overdensities \citep{cai:2016,nad+perc:2018}. On smaller
scales, nonlinear effects such as velocity dispersion may lead
to extended profiles along the line of sight \citep{cai:2016}, although the magnitude
of these effects is unclear \citep[see discussion in][]{nad+perc:2018}. We replicate these findings for simulated
voids at $z\sim2$ in the underlying flux and density fields for the entire $256^3$ $h^{-3}$ Mpc$^3$ box.

However,
we find that when measured in $(32 \times 32 \times 256)$ $h^{-3}$ Mpc$^3$ CLAMATO-like subvolumes,
the void quadrupole is significantly
distorted by edge effects in the Wiener filter
and void finder.  We also find that the void quadrupole is
significantly distorted by continuum error, since continuum error is correlated
along the line of sight.  Due to the large impact of these systematic effects, we do not
present redshift-space distortion measurements in CLAMATO voids here.  Future surveys with larger contiguous area
(e.g. an IGM tomography survey on the Subaru Prime Focus Spectrograph over 20 deg$^2$)
will be less impacted by continuum errors: we find very good agreement between
the void quadrupole in a $128 \times 128 \times 256$ $h^{-3}$ Mpc$^3$ subvolume and the full
TreePM box.  However, continuum error will remain a major source of systematic
error for modeling redshift-space distortions in Ly$\alpha$ forest voids: either the effects
of continuum error must be removed, e.g. by ignoring correlated pixels along the line of sight,
or we require accurate end-to-end modeling of the effects of continuum error on void shapes.
  
\section{Conclusions}
\label{sec:conclusions}

We present the first detection of cosmic voids at $z \sim 2$
using a spherical overdensity finder applied to a tomographic map
of the 3D Ly$\alpha$ absorption field from the CLAMATO survey carried out
on the Keck-I telescope.
By targeting background LBG and quasar sightlines
with mean transverse separation 2.5 \hMpc{} at 
$z \sim 2.3$, we create a Wiener-filtered map of the neutral hydrogen density
on few Mpc scales, which is an excellent tracer of the underlying
matter density.  This allows us to measure the density field on scales
considerably smaller than current galaxy surveys can achieve at this
redshift, enabling cosmic void detection at far greater ($\sim 1.7\times$)
cosmic distance than hitherto the most distant cosmic voids at $z\sim 1$.

Building on the results of \citet{stark:2015a},
we use realistic mock observations based on hydrodynamical and $N$-body simulations to 
calibrate thresholds for identifying voids in IGM maps. 
This is necessary to better model the Lyman-$\alpha$ forest
and continuum errors in the survey, which were neglected in \citet{stark:2015a}. Within the simulations, we find
worse void recovery from IGM tomography than \citet{stark:2015a}: $\sim 40\%$
of tomographically-identified voids are well-matched to density-field voids for $r \geq 5$ \hMpc{}.

Using thresholds calibrated from simulations, we apply the void finder
to the CLAMATO map to find a 19.5\% volume fraction of voids.  After removing
voids affected by edge effects,
we find good agreement between the void radius function in simulations and data.
Excursion set models can fit the void radius function only if
the excursion set threshold is adjusted considerably from the \citet{sheth_vdw:2004}
prediction of $-2.71$.

We also study the stacked void profiles for the higher-confidence subsample of 
large ($r \geq 5$ \hMpc{})
voids.  
As in \citet{stark:2015a},
we find no compensation ridge in the radial void profiles,
consistent with other spherical overdensity finders \citep{white_padmanabhan:2017}.

We validate the void detection by finding
that these voids are $\sim 6 \sigma$
underdense in coeval galaxies from the
MOSDEF, VUDS, and zCOSMOS spectroscopic redshift surveys, as well
as CLAMATO-confirmed galaxies falling within
the tomographic volume.  While the galaxy catalogs
are unable to detect voids on their own,
they validate the detection
of voids in IGM tomography
by showing that our voids have significantly
fewer galaxies than random regions with the same
radius distribution.

Identifying cosmic voids requires both a large
volume and a reasonably dense sampling of the 
density field.  Previous detections of voids
from galaxy surveys have extended to $z \sim 1$
\citep{conroy+05,cecc+06,sut+12,mich+14,mao+17,san+17},
while IGM tomography can detect voids at $z \sim 2.3$,
providing by far the most distant sample of voids
owing to much denser sampling of the density field
than galaxy surveys at comparable redshifts.
Moreover, upcoming surveys will dramatically
increase the number of $z \sim 2.5$ voids detected
via IGM tomography.  We find 48 voids with $r \geq 5$
\hMpc{} (for which we expect $\geq 45\%$ void recovery); 
the full CLAMATO survey will cover $\sim 3-5$
times more volume than the data used in this paper
and thus we expect to find $>100$ $r \geq 5$ \hMpc{} voids,
in line with the estimates in \citet{stark:2015a}.
Moreover, the Prime Focus Spectrograph (PFS) on the
Subaru telescope will
begin operation by 2020 \citep{takada_pfs}; it will
allow for surveys covering a much wider area, owing
to the much larger field of view of PFS compared to LRIS. 
A dedicated IGM tomography survey
building on the PFS galaxy evolution survey could
cover 15-20 deg$^2$ with sightline separation 3-4 \hMpc{}, 
i.e.\ comparable or slightly worse sampling than CLAMATO ---
the exact parameters are currently under discussion within the PFS collaboration.
Thus, such a survey on PFS could find 2000 $z \sim 2.5$ voids \citep{stark:2015a} with comparable fidelity to CLAMATO.  The larger area could be particularly crucial
to detecting void redshift space distortions at high significance.

At low redshifts, voids have been used for Alcock-Paczynski tests to measure cosmological parameters,
since voids are on average spherical in real-space \citep{sutter:2014,mao+17b}.
\citet{stark:2015a} estimates that a competitive high
redshift measurement of the Alcock-Paczynski parameter
will require 10,000 voids, which could be achieved
by a dedicated 100-night tomography survey on PFS, or by shorter surveys on even more ambitious instruments
such as the Maunakea Spectroscopic Explorer \citep{mcconnachie:2016} or the Billion Object Apparatus \citep{dod+16}.
On the other hand, \citet{stark:2015a} find that
linear theory accurately predicts
the radial velocity profile of voids, suggesting
that studying the velocity field either to infer cosmological parameters \citep[e.g.][using redshift-space distortions at low redshift]{ham+16} or
to test modified gravity theories
could be promising avenues of exploration.
In particular, \citet{clamp+13} estimates that modified
gravity theories could alter void profiles
in a way that could be observed with samples
of 20 voids.

Finally, voids offer an intriguing test-bed
for galaxy formation, as they contain
halos that have grown primarily by diffuse accretion
rather than mergers \citep{fakhouri+ma:2009}.
Existing studies of galaxy formation in voids
have been limited to low redshift,
where differences in void galaxy properties
can be attributed largely to their different
stellar masses \citep{hoyle+05,tinker:2008,alp15,penny:2015,beygu:2016}.  However, this may be
different at high redshift, particularly
since the global star formation rate
at $z \sim  2$ is much higher than at $z \sim 0$.  In principle, we have already
identified 35 galaxies in voids; however, if
a tomography-identified void contains a galaxy,
it is more likely that it is a fluctuation
due to noise than otherwise.  \citet{stark:2015a}
point out that ``true'' voids are expected
to be devoid of such bright galaxies,
but that voids could contain faint $L \sim 0.3 L_{\star}$ galaxies 
that could be observed
by the NIRSPEC spectrograph on JWST.

\acknowledgements{
We thank Shirley Ho, Uro\v{s} Seljak, Joanne Cohn, and Zachary Slepian
for helpful comments on this work.
 K.G.L. acknowledges support for this
work by NASA through Hubble Fellowship grant HF2-51361
awarded by the Space Telescope Science Institute, which is
operated by the Association of Universities for Research in
Astronomy, Inc., for NASA, under contract NAS5-26555.
  We are also grateful to the entire COSMOS collaboration for their assistance and helpful discussions.
  Calculations presented in this paper used resources of the National Energy Research Scientific Computing Center (NERSC), which is supported by the Office of Science of the U.S.~Department of Energy under Contract No.~DE-AC02-05CH11231.
The data presented herein were obtained at the W.M. Keck Observatory, 
which is operated as a scientific partnership among the California Institute of Technology, 
the University of California and the National Aeronautics and Space Administration (NASA). 
The Observatory was made possible by the generous financial support of the W.M. Keck Foundation.
The authors thank Yong's Kal-Bi in Waimea, HI, for vital sustenance during our observations.
  The authors also wish to recognize and acknowledge the very significant cultural role and reverence that the summit of Maunakea has always had within the indigenous Hawai'ian community.  We are most fortunate to have the opportunity to conduct observations from this mountain. 
  }
  
  \appendix

We have made the CLAMATO void catalog (Table~\ref{tab:all_voids}) publicly available on Zenodo (\url{https://doi.org/10.5281/zenodo.1295839}).
We have also included void catalogs from the mock CLAMATO-like observations in the Nyx and TreePM box,
including void catalogs from both the full TreePM box and the 64 subvolumes, and void catalogs from
the (real and redshift-space) density fields and underlying flux of the Nyx simulation, corresponding to the
void fractions reported in Table~\ref{tab:volume_fraction}.
We have also included the mock CLAMATO maps from these simulations.

\bibliographystyle{yahapj}
\bibliography{citations.bib}

\end{document}